\documentclass[]{elsarticle}
\usepackage[margin=1in]{geometry}
\usepackage{hyperref}
\usepackage{multirow}
\usepackage{array}
\usepackage{bm}
\newcolumntype{P}[1]{>{\centering\arraybackslash}p{#1}}
\newcolumntype{M}[1]{>{\centering\arraybackslash}m{#1}}
\usepackage{amssymb}

\journal{Elsevier}
\biboptions{authoryear}

\newcommand{\Lagr}{\mathcal{L}}

\begin{document}

\begin{frontmatter}

\title{AIFNet: Automatic Vascular Function Estimation for Perfusion Analysis Using Deep Learning
}
%\tnotetext[mytitlenote]{Fully documented templates are available in the elsarticle package on %\href{http://www.ctan.org/tex-archive/macros/latex/contrib/elsarticle}{CTAN}.}

%\author{Ezequiel de la Rosa, Diana M. Sima, Bjoern Menze, Jan S. Kirschke, David Robben } %\fnref{myfootnote}

\author[icoaddress,tumaddress]{Ezequiel de la Rosa\corref{mycorrespondingauthor}}
\cortext[mycorrespondingauthor]{Corresponding author}
\ead{ezequiel.delarosa@icometrix.com}

\author[icoaddress]{Diana M. Sima}
\author[tumaddress]{Bjoern Menze}
\author[tumaddress2]{Jan S. Kirschke}
\author[icoaddress,kuadress,kuadress2]{David Robben}
\address[icoaddress]{ico\textbf{metrix}, Leuven, Belgium}
\address[tumaddress]{Department of Computer Science, Technical University of Munich, Munich, Germany}
\address[tumaddress2]{Neuroradiology, School of Medicine, Technical University of Munich, Munich, Germany}
\address[kuadress]{Medical Imaging Research Center (MIRC), KU Leuven, Leuven, Belgium}
\address[kuadress2]{Medical Image Computing (MIC), ESAT-PSI, Department of Electrical Engineering, KU Leuven, Leuven, Belgium}

\begin{abstract}
%Acute ischemic stroke treatment relies on reperfusion techniques, which aim to reestablish the blood supply to the brain parenchyma and potentially recover tissue at risk. Whether to reperfuse or not is mainly decided based on the quantification of the salvegable $penumbra$ and irreversibly damaged $core$ lesions, which are identified through perfusion imaging techniques. %

Perfusion imaging is crucial in acute ischemic stroke for quantifying the salvageable $penumbra$ and irreversibly damaged $core$ lesions. As such, it helps clinicians to decide on the optimal reperfusion treatment. In perfusion CT imaging, deconvolution methods are used to obtain clinically interpretable perfusion parameters that allow identifying brain tissue abnormalities. Deconvolution methods require the selection of two reference vascular functions as inputs to the model: the arterial input function (AIF) and the venous output function, with the AIF as the most critical model input. When manually performed, the vascular function selection is time demanding, suffers from poor reproducibility and is subject to the professionals' experience. This leads to potentially unreliable quantification of the penumbra and core lesions and, hence, might harm the treatment decision process. In this work we automatize the perfusion analysis with AIFNet, a fully automatic and end-to-end trainable deep learning approach for estimating the vascular functions. Unlike previous methods using clustering or segmentation techniques to select vascular voxels, AIFNet is directly optimized at the vascular function estimation, which allows to better recognise the time-curve profiles. Validation on the public ISLES18 stroke database shows that AIFNet reaches inter-rater performance for the vascular function estimation and, subsequently, for the parameter maps and core lesion quantification obtained through deconvolution. We conclude that AIFNet has potential for clinical transfer and could be incorporated in perfusion deconvolution software.
\end{abstract}

\begin{keyword}
\texttt Ischemic stroke\sep Perfusion imaging \sep Arterial input function\sep Deep learning
\end{keyword}

\end{frontmatter}

\section{Introduction}
Stroke is currently the second leading cause of mortality and the third leading cause of disability worldwide \citep{trialists2013organised}. In physio-pathological terms, it is defined as an \textit{‘acute neurologic dysfunction of vascular origin with sudden (within seconds) or at least rapid (within hours) occurrence of symptoms and signs corresponding to involvement of focal areas in the brain’} \citep{force1989stroke}. Two main types of the disease can be recognised: ischemic and hemorrhagic, representing 85\% and 15\% of total cases respectively \citep{hinkle2007acute}. We focus on the ischemic case, where there is a shortage in the blood supply to the brain tissue, cutting the provision of oxygen and glucose. During the ischemic event, brain tissue might become necrotic (i.e., cells are dead and the tissue is irreversibly damaged, known as $core$) or in a hypo-perfused but salvageable state (i.e., tissue is at risk but could return to a healthy condition, known as $penumbra$) \citep{robben2016image}. %On the other hand, hemorrhagic stroke is caused by an artery rupture, which produces hypoperfusion of some brain territories and increases intracranial pressure \citep{robben2016image}.
\subsection{Perfusion CT in Acute Ischemic Stroke}
\begin{figure}[!t]
\hspace*{3cm} 
\includegraphics[scale=.4]{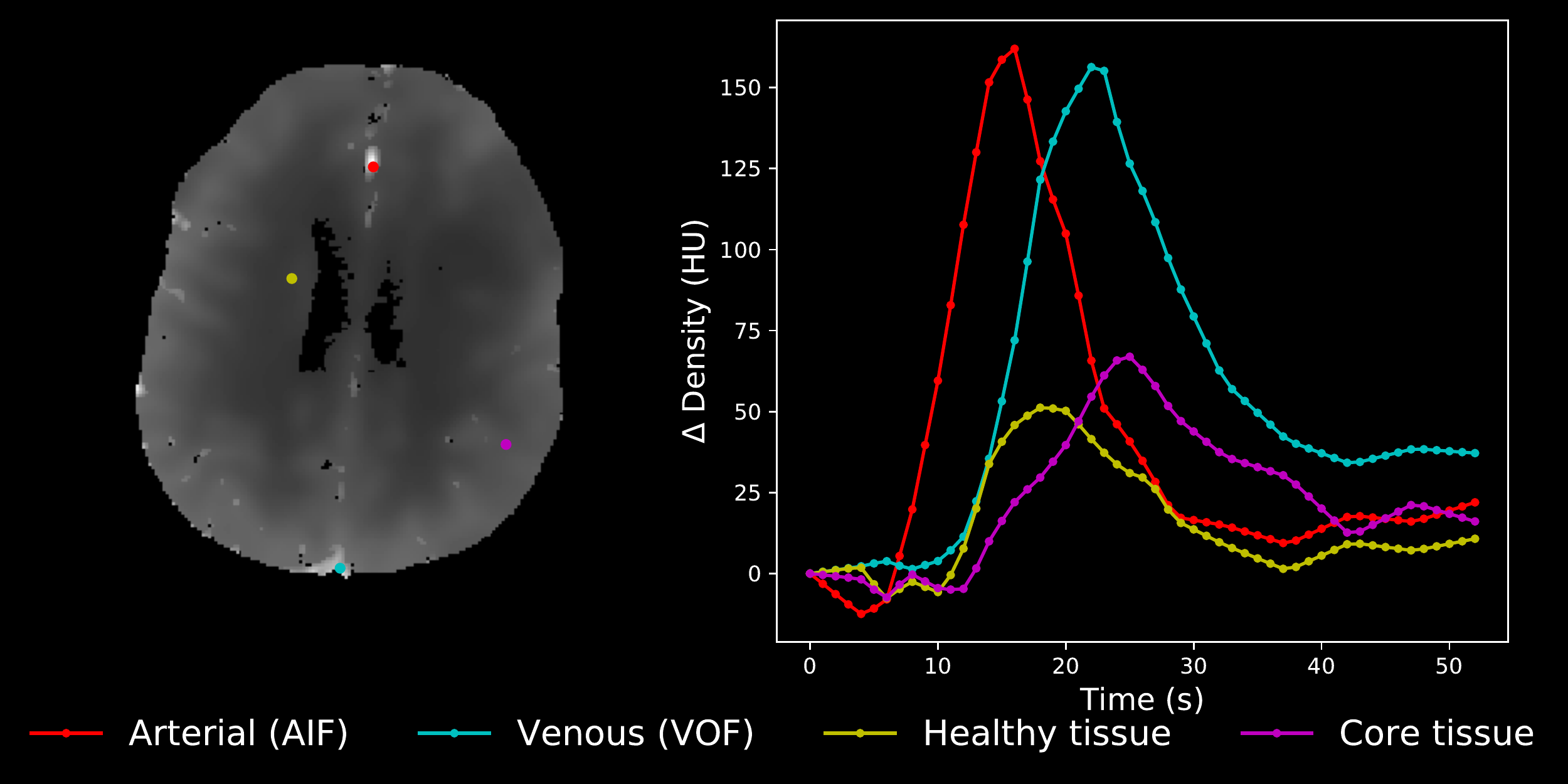}
\caption{Contrast enhancement curves for different brain tissues. Left: a perfusion CT example from ISLES18. Right: Corresponding time curves at the indicated locations. Healthy and diseased brain areas have been identified through diffusion weighted imaging. The healthy and core time-curves are scaled by a factor of six for visualization. HU: Hounsfield units.}
\label{fig:tissue_curves}
\end{figure}

Acute ischemic stroke therapies rely on reperfusion techniques, where the main goal is to reestablish the blood flow supply in the affected territories by thrombolysis or thrombectomy \citep{campbell2018imaging}. Identifying which patients might benefit from these treatments is critical for clinical decision making \citep{campbell2018imaging, albers2016ischemic}. To this end, assessment and quantification of the core and penumbra tissues are required. In the acute scenario, computer tomography (CT) is the most widely used imaging technique, where perfusion CT (CTP) enables the determination of the core and penumbra areas \citep{konstas2009theoretic}. An iodinated contrast agent is intra-venously injected in the patient for 7-10 s, and continuous CT acquisition is followed for around 50 s \citep{fieselmann2011deconvolution}. As such, 4D data is generated, resulting in a brain volume imaged during the agent passage through the brain vessels and parenchyma. The process for evaluating brain tissue status is performed by firstly obtaining parameter maps from the CTP time series and by later applying a tissue discrimination rule (mainly, thresholding). Typical maps include cerebral blood flow (CBF), cerebral blood volume (CBV), time to peak (TTP) and time to the maximum of the residue function (Tmax). It is worth saying that there is no $gold$ standard for quantifying perfusion metrics \citep{lorenz2006automated}, and all methods found in literature provide merely non-exact solutions. Experimental studies have shown that CBV and CBF discriminate ischemic and oligemic tissue with 90.6 \% and 93.3 \% sensitivity and specificity respectively when using histological measurements as ground truth \citep{murphy2007serial}. The most widely used methods for CTP parameter map estimation are based on deconvolution \citep{konstas2009theoretic}, which provide a solution to the indicator dilution theory described by:
\begin{equation}
    c_{tissue}(t) = c_{art}(t)\circledast h(t)
    \label{eq:dil_theory}
\end{equation}
where $c_{tissue}(t)$ represents the CTP contrast enhancement in a voxel of tissue, $c_{art}(t)$ is the contrast enhancement in the arteries (known as arterial input function, from now on `AIF'), $h(t)$ is the flow-scaled residue function and $\circledast$ represents the convolution operator. The delay-invariant singular value decomposition deconvolution is the preferred technique for algebraically solving Eq. \ref{eq:dil_theory} and it is widely implemented in software packages \citep{fieselmann2011deconvolution, konstas2009theoretic, kosior2007perftool, kudo2010differences,vagal2019automated}. The method has been extensively validated in clinical practice, showing better performance compared to similar techniques \citep{konstas2009theoretic, fieselmann2011deconvolution} like the maximum slope approach \citep{konstas2009theoretic, klotz1999perfusion}, non-delay invariant deconvolution \citep{ostergaard1996highI,ostergaard1996highII}, etc. Deconvolution methods require as input to the algorithm the CTP series and two vascular functions: the AIF and the venous output function (VOF). These vascular functions are reference time-curves representing the contrast concentration inlet and outlet to the tissue under consideration $c_{tissue}(t)$. Fig. \ref{fig:tissue_curves} shows an example case of vascular functions (i.e. AIF and VOF) and contrast enhancement curves for healthy and core tissue areas. In clinical practice the AIF and VOF are generally selected by a radiologist, a time demanding and highly variable process that implies selecting in the CTP series the optimal candidate voxels. Frequently, a single voxel per vascular function is selected, which leads to low SNR curves. Voxel selection is, moreover, subject to the professionals' training and experience, which not only introduces human bias \citep{lorenz2006automated} but it may also affect CBF maps depending which side of the brain the AIF is chosen from \citep{wu2003tracer, thijs2004influence}. The AIF is so critical for generating accurate maps that very small changes in its shape and/or location may produce a profound effect over the generated maps \citep{mlynash2005automated, mouridsen2006automatic}. Besides, given the acute context of the disease, a fast voxel selection has to be performed. It has been shown that for every 30-minute delay in reperfusion, the probability of good outcome decreases by 20\% \citep{khatri2014time}. Given these limitations, automatic, fast and reproducible core and penumbra quantification are highly desired.

\subsection{Automatic core and penumbra segmentation}
Automatic machine and deep learning approaches for core and penumbra quantification have been explored in two ways: 1) by direct parameter maps estimation and 2) by direct lesions segmentation. On one hand, automatic parameters maps estimation (i.e., bypassing deconvolution) was explored in 
\citep{mckinley2018machine,robben2018perfusion,ulas2018direct,ulas2018convolutional,meier2019neural}. However, the main drawback of these methods is the fact that $silver$ standard maps obtained through deconvolution or other methods (e.g. compartmental models in the case of perfusion MRI) are approximated. Note that in these approaches there is also an AIF assumption behind the parameter maps ground truth. As such, these methods do not improve the perfusion gold standard, but aim to reproduce it with a different model. On the other hand, direct lesion segmentation approaches
use native CTP data with or without perfusion maps as model inputs. Thus, the neural networks are used for finding a non-linear transformation from CTP and/or CBF, CBV, MTT and Tmax that estimates brain lesions. For instance, in \citep{bertels2018contra} and \citep{robben2020prediction} direct lesion segmentation is conducted by only using CTP images. While the former work exploits contralateral brain information into a U-Net based architecture, the latter work includes metadata and vascular functions into a multiresolution DeepMedic-based \citep{kamnitsas2017efficient} architecture. Other works include parameter maps obtained through deconvolution as inputs to the model \citep{clerigues2019acute, abulnaga2018ischemic, song2018integrated,wang2020automatic}. Similarly as in \cite{bertels2018contra}, \cite{clerigues2019acute} exploit brain symmetry information with U-nets.  \cite{song2018integrated} and \cite{wang2020automatic} propose, instead, to synthesize pseudo diffusion weighted imaging (DWI) data to improve core lesion segmentation. While deep learning based approaches showed good overall performance, their main limitation is the poor model’s explainability and lack of quality control. Since these fully `black-box' methods do not allow AIF or perfusion maps inspection, they  preclude physicians to recompute the parameter maps with a manually corrected AIF in clinically or technically challenging cases. As such, the clinical transferability potential of these models is limited. In this work we aim to automatize, instead, the well validated deconvolution process by the automatic selection of vascular functions. In this way, we avoid approximating parameters that can be directly estimated through a physical model while also preserving explainability and quality control in clinical settings.

\subsection{Automatic Vascular Function Selection}
Automatic vascular function selection has been explored for perfusion MRI in \citep{murase2001determination, mouridsen2006automatic, peruzzo2011automatic, shi2014automatic, shi2000normalized, yin2015automated, fan2019automatic, winder2020automatic}. These methods mainly rely on clustering techniques, where fuzzy c-means \citep{murase2001determination}, K-means \citep{mouridsen2006automatic}, hierarchical clustering \citep{peruzzo2011automatic}, gamma-variates based clustering \citep{rausch2000analysis} and affine propagation clustering \citep{shi2014automatic} were explored. Heuristic approaches have also been traditionally used, where some rules are defined for finding the best-matching curve, such as in \citep{mlynash2005automated, rempp1994quantification}. Other techniques use normalized cuts \citep{shi2000normalized, yin2015automated} and independent component analysis \citep{calamante2004defining}. 
Moreover, vascular function estimation using deep neural networks can be conducted through segmentation approaches aiming to detect arterial/venous voxels candidates. \cite{fan2019automatic} proposed a deep learning segmentation approach for delineating AIF candidates in perfusion MRI. The method uses two independently optimized 3D CNNs for conducting arterial tissue segmentation: one extracting spatial information in the $x$-$y$-$z$ axis, and another one extracting temporal-information in the $x$-$y$-$t$ axis (with $t$ representing the temporal domain). Afterwards, the networks' results are merged using a late-fusion support vector machine. More recently, \cite{winder2020automatic} proposed a binary output CNN for classifying arterial $vs$ non-arterial voxels in CTP and perfusion MRI. The AIF is then estimated by averaging the most probable arterial voxels. Though segmentation or classification methods can identify potential good curves, they have some limitations: i) They require complete manual annotation of all ``good-looking" voxel curves, which is very time demanding and ii) They could not always guarantee optimal AIF curve selection (for a possible definition of optimal AIF selection, see Methods \ref{section:Vascular Function Annotation}) since the algorithms are mainly optimized to perform selection based on spatial information rather than time profiles. Segmentation methods may lead, for instance, to the selection of delayed input functions, which produce large errors in some deconvolution algorithms \citep{calamante2000delay, wu2003effects}. In CTP imaging, however, vascular function selection is under-explored. Excepting the work of \cite{winder2020automatic} the few existing methods are mostly private and patented, such as in \citep{bammer2014automated, shi2014systems}. Besides, most of the methods developed for perfusion MRI have not been validated for CTP. Despite perfusion CT and perfusion MRI having common working points, there are still technical differences that may affect the automatic selection of CTP vascular functions (such as lower tissue-density contrasts and lower SNR of CT compared to MRI). Moreover, additional challenges in CTP include overlapping density distribution of bone, artifacts and calcifications with the iodine contrast.

In this work we propose AIFNet, an end-to-end supervised convolutional neural network devised for estimating vascular functions (i.e. AIF and VOF) in perfusion imaging. The model is easy to train and deploy given the minimal data annotation required, which can be as little as a single voxel per vascular function. AIFNet receives 4D CTP series as input and generates as output i) the estimated AIF and VOF curves and ii) a voxel-wise, interpretable probability map representing the voxelwise contribution to the estimated vascular signal. Unlike other approaches, AIFNet is optimized at a vascular function level, which helps the network to better learn the time-curve profiles. The method preserves clinical interpretability and also enables quality control of the selected AIF/VOF brain vasculature, thus enhancing its clinical transferability potential. Through an extensive analysis at signal, parameter maps and lesion quantification levels, we show that our method performs as good as manual raters on the open ISLES18 acute stroke database.

\section{Methods}

\subsection{Function Estimation with Deep Learning}
AIFNet is a fully end-to-end deep learning approach for vascular function estimation. It works by estimating a 3D probabilistic volume that represents the voxelwise contribution to the vascular signal. The advantage of having an averaged curve using multiple voxels lies on the higher function’s SNR as well as on the method robustness. The network receives as input the 4D perfusion series $x(t)$ and outputs the predicted arterial and venous functions as $\hat{y}(t) = AIF_{Net} (x(t))$, being $x(t) = \{x_t; t = 1, 2, …, T\}$ with $x_t$ representing the sampled time point volumes of dimension $M$$\times$$N$$\times$$Q$. We want to find for the considered volume, its corresponding vascular functions (AIF and VOF, for simplicity not differentiated in the notation) represented by $\hat{y}(t)=\{\hat{y_t}; t = 1, 2, ..., T\}$, where $\hat{y_t}$ is the estimated signal at time $t$ (in Hounsfield units). For finding $\hat{y}(t)$, we represent each time point $\hat{y_t}$ as a weighted average of all voxels of the volume $x_t$ at that $t$ time point as:

\begin{equation}
\hat{y_t} =\sum_{q=1}^{Q}\sum_{n=1}^{N}\sum_{m=1}^{M} x_{t}(m,n,q)*P_{vol}(m,n,q)
\label{eqn:ypred}
\end{equation}
where $P_{vol}$ is the 3D probabilistic volume containing the voxelwise contribution to the vascular function and fulfilling:

\begin{equation}
\sum_{q=1}^{Q}\sum_{n=1}^{N}\sum_{m=1}^{M} P_{vol}(m,n,q)=1
\end{equation}

Our problem is hence confined to finding $P_{vol}$. With this aim, AIFNet receives as input native CTP series, and generates as outputs $P_{vol}$ and its associated vascular function. For the proper finding of $\hat{y_t}$ it is of paramount importance to penalize the predictions over time rather than in amplitude. This is due to two facts: $1$) perfusion analysis is independent of the AIF scaling (given the high partial volume effect in the arteries and the presence of confounders that affect the AIF, a recalibration step is always required for generating absolute parameter maps) and $2)$ a suboptimal deconvolution might occur by selecting delayed input functions. Later, for obtaining absolute parameter maps, a recalibration of the VOF is conducted prior to deconvolving the native CT volumes, as it will be explained in section \ref{section: Recalibration}. The penalty in the time domain is introduced by using Pearson’s correlation as loss function as follows:

%add equation 1
\begin{equation}
\Lagr(y(t), \hat{y}(t)) = -\frac{\sum_{t = 1}^{T}(y_t -\overline{y} )(\hat{y_{t}} -\overline{\hat{y}} )}{ \sqrt{\sum_{t=1}^{T}(y_t -\overline{y} )^2}\sqrt{\sum_{t=1}^{T}( \hat{y_{t}} - \overline{\hat{y}} )^2}} 
\end{equation}
where $y(t)$ and $\hat{y}(t)$ are the ground truth and predicted vascular functions with respective mean values $\overline{y}=\frac{1}{T}\sum_{t = 1}^{T}y_{t}$ and $\overline{\hat{y}}=\frac{1}{T}\sum_{t = 1}^{T}\hat{y_{t}}$. 

\subsection{Architecture}
\begin{figure*}[t]
\hfill\includegraphics[scale=0.6]{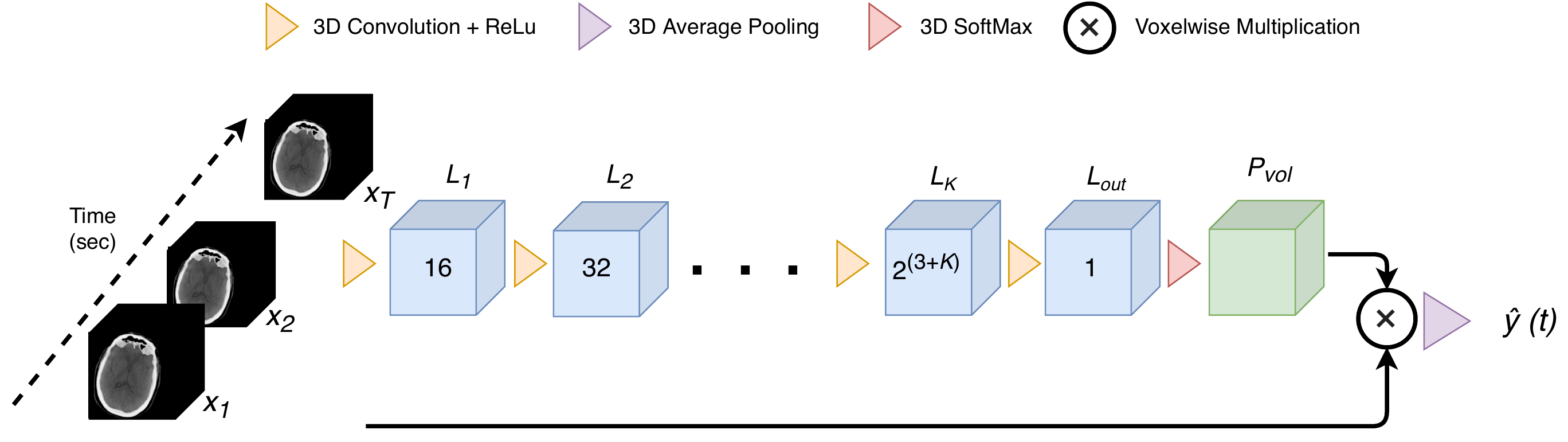}
\caption{AIFNet architecture. The CTP time points $x_t$ ($t = 1, 2, ..., T$) are incorporated as channels in the network. All convolutional layers use 3x3x3 kernels except $L_1$, which uses 3x3x1. $L_k$ is the $k-th$ convolutional layer (with $k = 1, 2, ..., K$). Inside each feature block the number of channels used is indicated. $P_{vol}$ is the probabilistic volume. The 3D average pooling block averages the volumetric information along the $x$-$y$-$z$ axes, such that the predicted vascular function $\hat{y}(t)$ is a 1D vector of length $T$. }
\label{fig:archi}
\end{figure*}

AIFNet architecture is shown in Fig. \ref{fig:archi}. It used 3D convolutional layers for volumetric feature extraction, which are finally translated into a probabilistic volume through a 3D softmax operation. After finding $P_{vol}$, a voxelwise multiplication and 3D average pooling blocks are used for obtaining $\hat{y}(t)$, by means of equation \ref{eqn:ypred}. Each convolutional layer $L_k = \{k = 1, 2, …, K\}$ has $2^{3+k}$ filters with a 3x3x3 kernel with exception of $L_1$, which uses a 3x3x1 one with the aim of compensating the lower image resolution along the $z$-axis. The CTP time points are incorporated as channel information into the network. A fixed number of $T$ time points are used for all scans. In our experiments we use a $T$ equal to the smallest number of time points found among all scans.  Rectified linear units are used as activation functions \citep{krizhevsky2012imagenet}. For mapping the convolutional layers to a single probabilistic volume, we add an extra convolution block ($L_{out}$) with only one filter in between $L_K$ and the softmax operator. 

\subsection{Training phase}
\label{sec:Training phase}
The network is optimized using stochastic gradient descent with momentum. A batch size of one sample is used. Regularization of the model is reached using a perfusion-specific data augmentation approach.

\subsubsection{Perfusion Specific Data Augmentation} \label{section: data augmentation}
We adapt the data augmentation strategy proposed in \citep{robben2018perfusion} for working at an image level. Two perfusion specific phenomena are modelled: $i)$ the variability of the contrast bolus arrival, which depends on the injection protocol and the patient's cardiovascular system and $ii)$ the variability of the curve's peak-to-baseline (PTB) values, which depends on the iodine concentration in the contrast agent. Bolus arrival changes are simulated by randomly shifting the time attenuation curves, for which the first or last CT volumes are replicated (late or early simulated arrivals respectively). On the other hand, curve PTB changes are simulated in a three-step approach. Firstly, the pre-contrast averaged volume is subtracted from the perfusion series. Secondly, a random scaling is applied. Thirdly, the pre-contrast volume is re-added to the perfusion series. Uniform distributions are used for simulating the random time shifts and the random PTB scaling. 

\subsection{Testing phase}
In the testing scenario, vascular function predictions are obtained by feeding the parametrized AIFNet model with the unseen CTP scans. The voxelwise multiplication and 3D average pooling blocks of AIFNet are performed over the full-length CTP perfusion series, with the aim of obtaining vascular function predictions that preserve the same number of time points as the native CTP scan. For VOF a signal recalibration step is also applied, as detailed below. 

\subsubsection{VOF Signal Recalibration} \label{section: Recalibration}
Our multiple signal averaging approach has the disadvantage of underestimating the VOF peaks. Since the VOF’s role in deconvolution-based perfusion analysis is to compensate for partial volume effect in the AIF by its recalibration, it is important that its PTB matches the same amplitude as single CTP candidate voxels. Ideally, a suitable VOF curve has the highest PTB value among all voxel candidates. Therefore, we use a probabilistic volume that encodes voxelwise contribution to the function estimation. Firstly, we generate a 3D volume encoding the voxelwise PTB values. Secondly, we scale this volume with $P_{vol}$ in order to obtain probabilistic-weighted PTB values. The VOF is finally recalibrated with the maximal value found in the weighted PTB distribution. We prefer using weighted PTB instead of only considering $P_{vol}$, since the highest probability voxel of $P_{vol}$ might have a low PTB, thus leading to an underestimation of the VOF PTB value.

\section{Experiments}
\subsection{Data}
\subsubsection{ISLES18}
The large public multi-center and multi-scanner ISLES18 dataset is used for our experiments \citep{maier2017isles, kistler2013virtual, cereda2016benchmarking}.  It consists of 156 CTP acquisitions acquired from 103 acute stroke patients from three US centers and one Australian center. In the ISLES challenge, data is split into a train (94 CTP volumes scanned from 63 patients) and a test (62 CTP volumes scanned from 40 patients) sets. The mismatch between patients and scans is due to the limited field of view of some scanners, which leads to two independent CTP acquisitions from different brain regions in some cases. We have directly accessed the clean and preprocessed data through the ISLES challenge site (http://www.isles-challenge.org/). For each acquisition, CTP and DWI data were performed within 3 hours of each other. The open database provides CTP scans for the whole dataset and infarct core lesion masks (delineated in DWI images) for the training set only. Subjects having more than 50\% of the DWI lesion with normal perfusion at the moment of the CTP acquisition were excluded, as well as those subjects with bad quality of the baseline CTP data and/or with inappropriate image coregistration due to distortions \citep{cereda2016benchmarking}. CTP volumes have been motion corrected and coregistered for matching the DWI lesion masks. Finally, scans have been spatio-temporally resampled (with a 256x256 dimension matrix and with a temporal resolution of one volume per second).  For a more detailed description of this database the reader is referred to \citep{cereda2016benchmarking}.

\subsubsection{Vascular Function Annotation} \label{section:Vascular Function Annotation}
All training and testing scans are in-house annotated by two independent raters (DR $\&$ EdlR). A single \textit{global} AIF and VOF per scan is selected (i.e., functions are measured from a major artery/vein and used as global inputs for the tissue in the whole brain \citep{calamante2013arterial}), where the following AIF time attenuation curves are preferred: $i)$ contralateral voxels to the affected area (rather than ipsilateral ones) \citep{kealey2004user, calamante2013arterial} , $ii$) Early bolus arrival AIF curves with a large and narrow peak enhancement \citep{calamante2013arterial} $iii)$ Curves with high contrast-to-noise ratio and, ideally, less affected by partial volume effect (qualitatively assessed) \citep{calamante2013arterial}. The $best$ voxel candidate (following the just mentioned criterion) among the anterior cerebral arteries, middle cerebral arteries, internal carotid arteries or the basilar artery are chosen as AIF. On the other hand, VOF curves are located in the superior sagittal, transverse or sigmoid sinuses, since being large vessels are less affected by partial volume effect than other vessels. All vascular function annotations are provided as supplementary material.  

\subsection{Performance assessment}
In order to evaluate the performance of AIFNet, we conduct a 5-fold (train 70\%, validation 10\%, test 20\%) cross-validation experiment using the annotations of rater \#1. All training and testing cases of the ISLES18 database are used in this experiment. In each fold, the train set is used to parametrize the network, the validation set to apply an early-stopping criterion (with the aim of avoiding overfitting) and the test set is independently used for predicting unseen cases. For an in-depth evaluation of the proposed method, results are assessed at a signal, parametric map and lesion quantification level.

\subsubsection{Vascular Function}
\begin{figure}[!t]
\hspace*{3.5cm} 
\includegraphics[scale=.6]{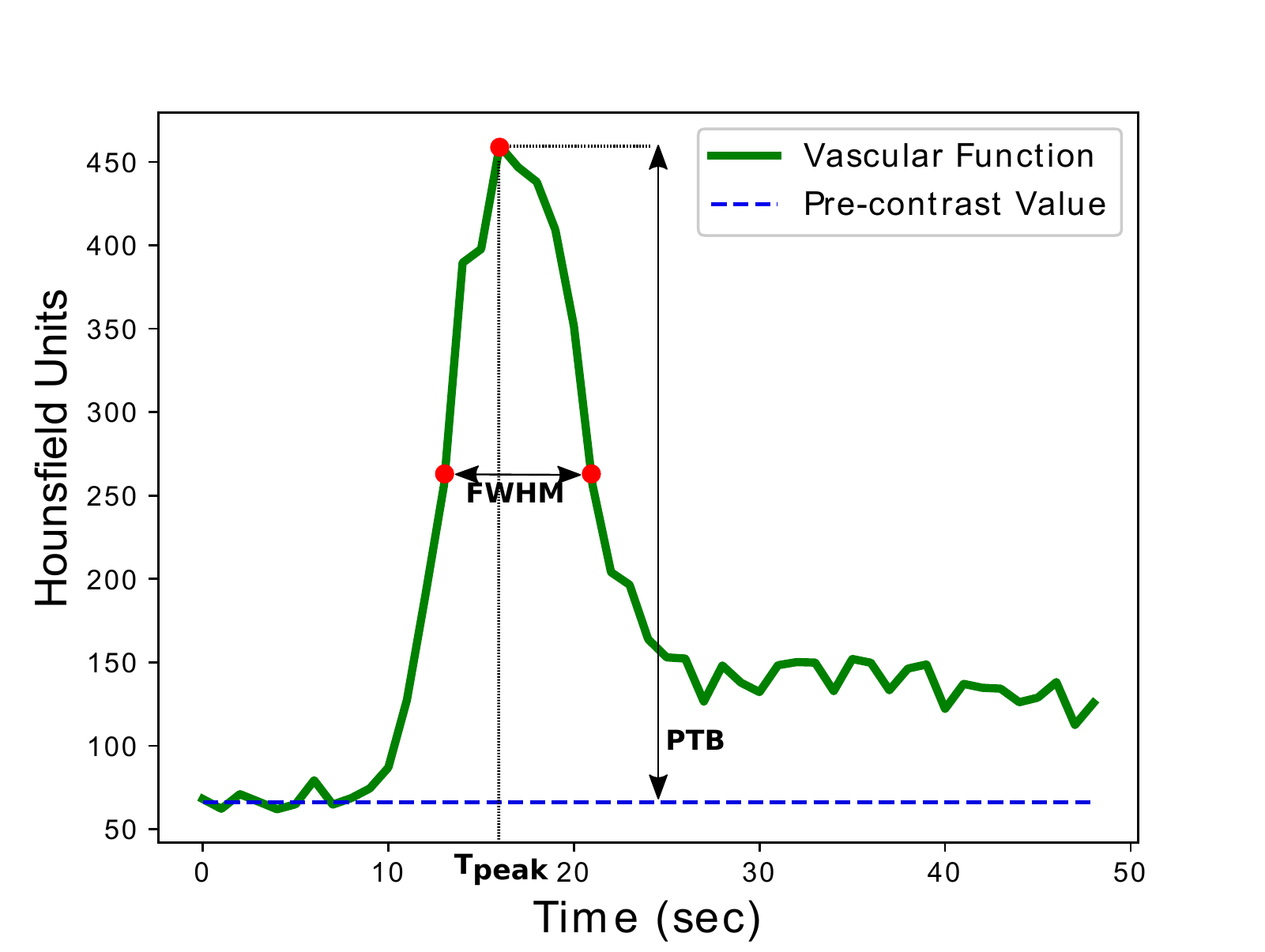}
\caption{Vascular function metrics. FWHM: Full width at half maximum; PTB: Peak to baseline; $T_{peak}$: Time at which the curve peak occurs.}
\label{fig:signal_metrics}
\end{figure}

Since there is no ground truth for the vascular functions we compare our predictions $\hat{y}(t)$ against the manual annotations $y(t)$ of both the raters (from now on, we refer to them as $y_{r1}(t)$ and $y_{r2}(t)$ for rater $\#$1 and rater $\#$2 respectively). The agreement between $y(t)$ and $\hat{y}(t)$ is computed over the time domain since perfusion analysis is independent of the AIF scaling (see sections \ref{sec:Training phase} and \ref{section: Recalibration}). To this end, we measure the time at which the curve peak occurs (namely $T_{peak}$), which should indicate potential time shifts of the predictions with respect to ground truth. Moreover, as a measure of the function’s width, we quantify the full-width at half-maximum (FWHM) interval. FWHM points are preferred to the curve’s onset/offset since these are more difficult to measure and in many cases the offset point is missing. Besides, for evaluating the VOF recalibration strategy, we assess the whole signal agreement using mean squared error and the measured PTB values. All signal metrics are illustrated in Fig. \ref{fig:signal_metrics}.

\subsubsection{Parameter Maps and Lesion Quantification}
Parameter maps (CBF, CBV, MTT, Tmax) are computed using the well known time-delay invariant singular value decomposition deconvolution. The method is the most commonly found in (clinical) software. Since deconvolution is a mathematically ill-conditioned problem, regularization techniques are necessary. We use Tikhonov regularization over the singular values under a Volterra discretization scheme \citep{sourbron2007deconvolution}. Absolute and relative parameter maps are computed, where the relative ones are calculated by the voxelwise normalization of the absolute ones with the mean control tissue region value. Control tissue is defined as the region with normal perfusion (i.e., $T_{max} < 6$ s \citep{lin2016whole}). 

To understand the impact of the vascular functions on the perfusion metrics we compare the parameter maps (obtained through deconvolution) between the automatically and manually annotated vascular functions. The same, unaltered deconvolution strategy is adopted in both cases. In this way we are sure that variations are only due to the vascular functions. However, the assessment of the method in lesion quantification terms is conducted by comparing the obtained segmentations against the ISLES18 ground truth lesion masks. Note that this comparison is only done for the ISLES18 training set, since lesion masks are not available for the test set. For segmenting the core lesions we use the rCBF map (obtained with the manually and automatically estimated vascular functions) thresholded at 38\%, a cutoff that was found earlier to be optimal \citep{cereda2016benchmarking}.  

\subsubsection{Ablation Analysis and Comparison with other CNNs}
We ablate our network for finding the optimal architecture and training strategy for computing $\hat{y}(t)$. The ablation is conducted for the AIF since it is the most critical input to the deconvolution model and it is much more difficult to estimate than the VOF. Besides, our method is compared against two similar CNNs: a regression and a segmentation network,  both of them modified versions of AIFNet. For having comparable approaches, we keep the network's architecture and configuration as close as possible to AIFNet. The same perfusion-specific data augmentation of section \ref{section: data augmentation} is used in all cases. Optimization is conducted using sthochastic gradient descent with momentum with an unitary batch size for both approaches. All the experiments are conducted using the original train-test data split of the ISLES18 challenge. For training purposes we randomly exclude 10\% of the train data and use it as validation set, assuring that in all experiments the same train-validation-test sets are used.

\paragraph{Regression AIFNet}
It consists of a 3D + 2D neural network equipped with six convolutional layers with average pooling and with two fully connected layers at the end. The last fully connected layer is a 1D vector with same number of neurons as time points in the perfusion CTP and represents the vascular function prediction $\hat{y}(t)$. The 3D to 2D data transformation in the network is conducted by squeezing the $z$-axis information by means of average pooling. Homogenizing the $z$-axis dimension is required for dealing with the variable CTP coverage (varying between 2 and 16 slices per scan). The 3D convolution kernels have dimension 3x3x3 and the 2D convolution ones have dimension 3x3. The Pearson's correlation coefficient is preserved as loss function. 
 
\paragraph{Segmentation AIFNet}
This network is fed with AIF binary masks as ground truth. It is similar to AIFNet by preserving the whole architecture except the voxelwise multiplication and 3D average pooling blocks (see Fig. \ref{fig:archi}). Besides, the last convolutional block ($L_{out}$) has two kernels followed by a softmax operation for conducting background and foreground segmentation. For compensating the large class imbalance, this network is trained using weighted categorical cross-entropy as loss function. The AIF is then estimated as the average function among the top ranked voxels, for which the outperforming threshold value in terms of AIF Pearson's correlation is used.

\subsubsection{Statistical Analysis}
For evaluating the entire vascular signal and its metrics (i.e. $T_{peak}$, FWHM and PTB) correlation analysis are performed. Unless specified, we use Pearson correlation coefficients. Only in cases where outliers are present a Spearman correlation is preferred. Mean, standard deviation and (5th, 95th) percentiles are provided. Additionally, to assess a potential bias of the different metrics, we compute the mean and standard deviation of the errors.

The discriminant power of the different parameter maps in separating penumbra and necrotic tissue are evaluated in the hypoperfused brain region by a receiver operating characteristic curve (ROC). The ROC is obtained by thresholding the different maps themselves at different cutoff values on the entire dataset. The area under the curve (AUC)  is used as a general performance metric.

The assessment of the core lesions is conducted by comparing the segmentations obtained using the vascular functions obtained by the raters and by AIFNet against the ground truth masks delineated over DWI. The mean volume error, mean absolute volume error and volume correlation are used for quantifying lesion volumetric (ml) agreement with the ground truth. Dice coefficient is used as a segmentation performance metric. In all cases, unless specified, paired t-tests are performed after visual inspection of the data distributions. Under the presence of non-normal distributions, outliers, or heteroskedasticity, a paired Wilcoxon-test is preferred. The significance level is set in all cases to $\alpha$ = 0.05. 

\section{Results and discussion}
All models are trained on a machine with a Tesla K80 Nvidia GPU (12 Gb dedicated), with 64 gb RAM and an Intel Xeon E5-2686 v4 multiprocessor. The training stage takes $\sim$11 hours for an AIF/VOF model. Manual annotations take between 2 and 4 minutes for both functions per scan, depending on the number of slices of the volume. On the other hand, predictions take $\sim$6 seconds per each vascular function per scan.  

\subsection{Signal Agreement}
\subsubsection{AIF}
Table \ref{tab_aif} shows a summary of the performance of our method compared to the two raters. The automatic predictions obtain high agreement with both raters in all the metrics considered. It is noticeable that an overall better agreement with rater $\#$2 ($y_{r2}(t)$) is obtained, even when the network is trained using annotations from rater $\#$1 ($y_{r1}(t)$), suggesting good generalization at inter-rater level. When the entire vascular signal is evaluated, the method obtains Pearson’s $r$ values reaching the raters range. A slightly lower 5\% percentile is observed in the agreement between $\hat{y}(t)$ and $y_{r1}(t)$ when compared with the inter-rater agreement. This discordance is, however, not found when comparing $\hat{y}(t)$ with $y_{r2}(t)$, which obtains fully overlapping ranges with the inter-rater performance. The 95\% percentile obtained between AIFNet and the raters is, as expected, close to r = 1 but never reaching perfect agreement, due to the weighted multivoxel selection strategy proposed. 

When the method performance is assessed in terms of $T_{peak}$, a high correlation with the manual annotations is found. It can be observed that the $T_{peak}$ annotations of $y_{r2}(t)$ are slightly delayed when compared with the $y_{r1}(t)$ annotations, without presenting statistically significant differences (p-value = 0.18). The AIF functions that AIFNet selects are on average $\sim$ 0.5 seconds delayed when compared with the raters (p-value $<$ 0.01 when compared against both raters). This temporal trend toward delayed events explains the slight overall lower agreement between AIFNet and both raters. Similarly, the agreement that is obtained for the FWHM between AIFNet and the raters is slightly lower than the inter-raters level (p-values $<$ 0.01 when compared against both raters, and p-value = 0.88 between raters, Wilcoxon test), with similar trends towards both raters. The predicted FWHM windows are on average $\sim$ 1 second longer than the manual ones. These time differences found in $T_{peak}$ and FWHM with our method are below the temporal CTP resolution (one frame, the minimal possible). The main reason behind these differences is the flip side of the coin of the multivoxel selection strategy. Thus, vascular function estimation based on multiple voxels could not provide the $earliest$ bolus arrival with the $highest$ and $narrowest$ curves, but averaged values over the activated voxels. Selecting vascular functions with these characteristics are, hence, not always possible with our strategy, since generally a single or just a few voxels fulfill these requirements for AIF.

\begin{table*}[!t]
\caption{\label{tab_aif} AIF agreement among methods. Mean (standard deviation) provided. Note that the AIF agreement is measured over the time domain only, since a posterior signal recalibration using the VOF is required due to the high partial volume effect in the arteries. $y_{r1}(t)$: annotated vascular function of rater 1; $y_{r2}(t)$: annotated vascular function of  rater 2; $\hat{y}(t)$: prediction with AIFNet; r: Pearson's correlation coefficient; $T_{peak}$: time at which the peak of the curve occurs; FWHM: full-width at half-maximum.
}
\centering
\begin{tabular}{M{2cm}M{3cm}M{3cm}M{3cm}M{3cm}}
       &                 &               &              &               \\
\multicolumn{2}{c}{\textbf{Metric}} &  $\bm{y_{r1}(t)}$ \textbf{vs} $\bm{y_{r2}(t)}$   & $\bm{y_{r1}(t)}$ \textbf{vs} $\bm{\hat{y}(t)}$ & $\bm{y_{r2}(t)}$ \textbf{vs} $\bm{\hat{y}(t)}$\\
 \hline
\multirow{2}{*}{Signal }&  r     & 0.971 (0.075) & 0.965 (0.05) & 0.969 (0.04)  \\ 

       & r (5th, 95th Perc)    & (0.883, 1)       & (0.838, 0.997)  & (0.884, 0.997)   \\
\hline
\multirow{2}{*}{$T_{peak}$}   & r               & 0.964         & 0.94         & 0.951         \\
       & Error (s)  & -0.14 (1.29)  & -0.55 (1.75) & -0.41 (1.6)   \\
\hline
\multirow{2}{*}{FWHM }  & r               & 0.902         & 0.854        & 0.853         \\        & Error (s)  & -0.08 (1.74)  & -0.89 (2.14) & -0.81 (2.14)\\ 
\hline

\end{tabular}
\end{table*}

\begin{table*}[!b]
\caption{\label{tab_vof} VOF agreement among methods. Mean (standard deviation) provided. $y_{r1}(t)$: annotated vascular function of rater 1; $y_{r2}(t)$: annotated vascular function of  rater 2; $\hat{y}(t)$: prediction with AIFNet; r: Pearson's correlation coefficient; MSE: Mean squared error; HU: Hounsfield units; $T_{peak}$: time at which the peak of the curve occurs; FWHM: full-width at half-maximum. PTB: Peak-to-baseline; Perc: Percentile.
}
\centering
\begin{tabular}{M{1cm}M{3.4cm}M{3.2cm}M{3.2cm}M{3.2cm}}
       &                 &                 &                &                 \\
\multicolumn{2}{c}{\textbf{Metric}} &  $\bm{y_{r1}(t)}$ \textbf{vs} $\bm{y_{r2}(t)}$   & $\bm{y_{r1}(t)}$ \textbf{vs} $\bm{\hat{y}(t)}$ & $\bm{y_{r2}(t)}$ \textbf{vs} $\bm{\hat{y}(t)}$ \\
 \hline
\multirow{4}{*}{Signal }   & r     & 0.985 (0.047)   & 0.981 (0.069)  & 0.983 (0.051)   \\ 
       & r (5th, 95th Perc)    & (0.944, 1)         & (0.914, 0.999)    & (0.925, 0.999)\\
       \cline{2-5}
       
          & MSE (HU)     & 1424.050 (3622.420)   &  1234.577 (2623.260)  & 1558.214 (3740.499)   \\ 
       & MSE (5th, 95th Perc)    & (0, 7148.105)         & (16.552, 7024.535)    & (24.979, 8213.376)\\
       
\hline
\multirow{2}{*}{$T_{peak}$}   & r               & 0.98            & 0.955          & 0.963           \\
       & Error (s)  & 0.269 (1.145)   & -0.070 (1.69)  & -0.33 (1.51)    \\
\hline
\multirow{2}{*}{FWHM}    & r               & 0.829           & 0.827          & 0.911           \\
       & Error (s)  & 0.115 (2.276)   & -0.038 (2.420) & -0.154 (1.744)  \\
\hline
\multirow{2}{*}{PTB}   & r               & 0.921           & 0.953          & 0.919           \\
       & Error (HU)  & 10.999 (55.193) & 8.977 (44.452) & -2.022 (57.790)\\
\hline

% Old recal method  &  Bias (HU)  & 10.999 (55.193) & 2.301 (87.653) & -8.698 (96.164)
\end{tabular}
\end{table*}

\subsubsection{VOF}
In Table \ref{tab_vof} a summary of the performance of our method for VOF estimation is shown. A high agreement with the manual annotations is obtained, which is better than the performance obtained for AIF estimation. These results can be expected since VOF compared to AIF is less affected by partial volume effect, has higher SNR and hence provides lower inter-rater variability (Table \ref{tab_vof}). 

\begin{figure*}[!b]
\hspace*{-5cm} 
\hfill\includegraphics[width=\textwidth]{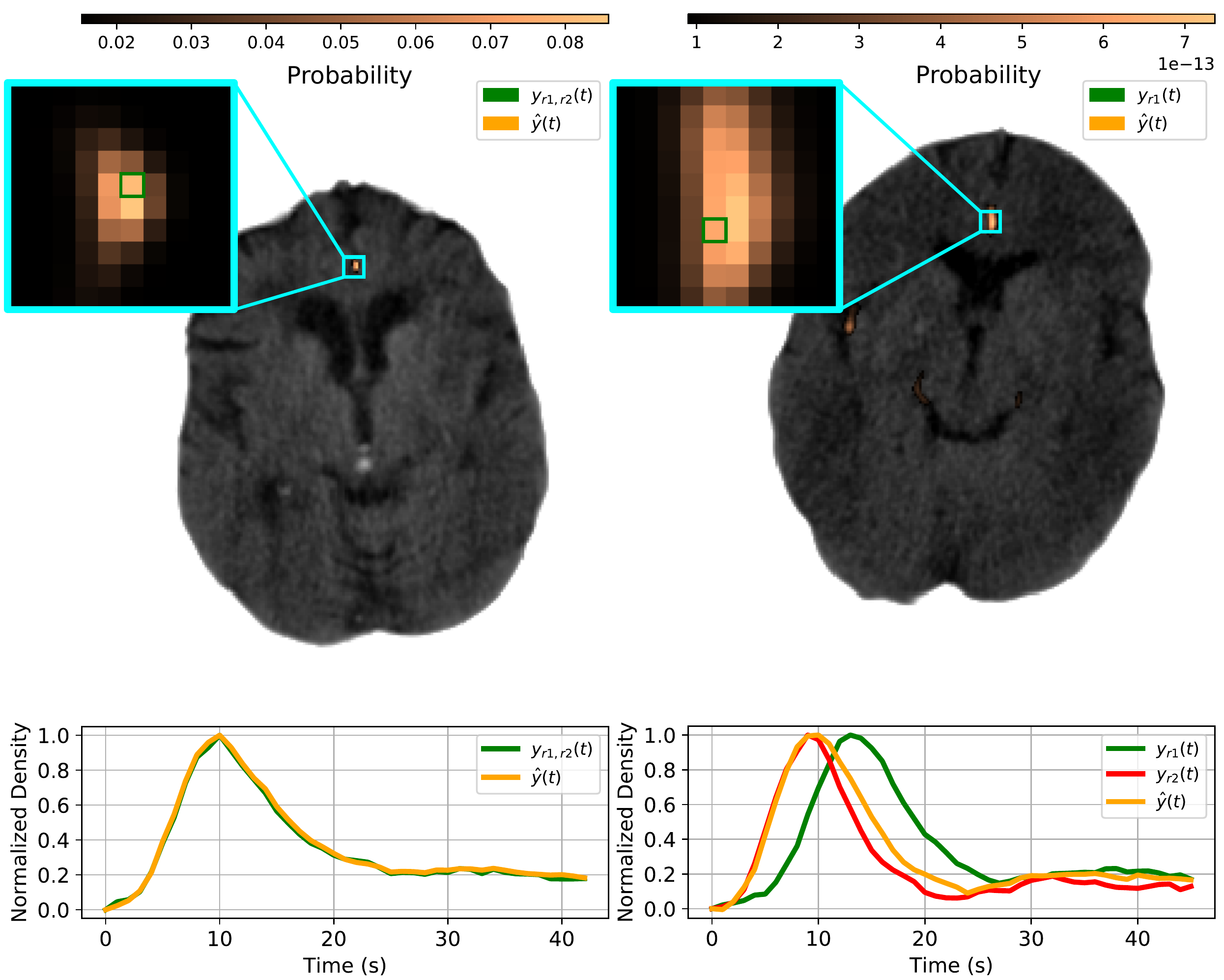}
\caption{Best (left) and worst (right) prediction performance in terms of Pearson's correlation between AIF functions. Above, the voxels selected by rater \#1 ($y_{r1}(t)$), rater \#2 ($y_{r2}(t)$) and AIFNet ($\hat{y}(t)$) as arterial input functions. Below, their corresponding vascular function. Note that in the best performance case, both raters have chosen the exact same voxel as AIF. In the worst performance case, the selected voxel location for $y_{r2}(t)$ is not shown since was annotated in a different volume slice. }
\label{fig:localization}
\end{figure*}

When the entire VOF signals are considered, a high correlation with the manual annotations is achieved, reaching inter-rater variability ranges. In terms of $T_{peak}$, a good performance is obtained. While there are statistically significant differences between the raters for this metric (p-value $<$ 0.01), the comparison between $y_{r1}(t)$ and $\hat{y}(t)$ does not show statistically significant differences (p-value = 0.61) while it does show for $y_{r2}(t)$ against $\hat{y}(t)$ (p-value $<$ 0.01). Moreover, the same delaying effect previously described for AIF is found, but in this case the delays are in the inter-rater range. For FWHM, the agreement between our method and rater $\#$2 is much higher than among raters. No statistically significant differences are found between the raters (p-value = 0.32) and between our predictions and the raters (p-value = 0.29 and p-value = 0.37 when comparing $\hat{y}(t)$ against rater $y_{r1}(t)$ and $y_{r2}(t
)$ respectively, Wilcoxon test). Finally, we evaluate the performance of our recalibration strategy by assessing the mean squared error between annotated and predicted signals, as well as between their corresponding PTB values. A high agreement and high correlation with manual annotations reaching inter-rater ranges are obtained for both metrics. In the assessment of the mean squared error, the inter-rater's 5th percentile is zero, which implies that the raters have sometimes selected the exact same voxel. The PTB comparison between $y_{r1}(t)$ and $y_{r2}(t)$ and between $y_{r1}(t)$ and $\hat{y}(t)$ shows statistically significant differences (p-value = 0.03 and p-value $<$ 0.01 respectively, Wilcoxon test) while the comparison between $y_{r2}(t)$ and $\hat{y}(t)$ does not (p-value = 0.051, Wilcoxon test). There is high agreement with the raters also in terms o PTB error, showing no clear trend of our method towards under/over-estimation of the VOF signals. 

\subsubsection{Arterial Localization} \label{section:localization}
The anatomical localization that AIFNet conducts can be assessed from the voxelwise activation encoded in $P_{vol}$. Unlike most AIF selection approaches selecting only few candidates, AIFNet allows multiple voxel contribution for building the vascular functions.

In Fig. \ref{fig:localization} the best and worst AIF (in correlations terms) among all predictions are shown. While the prediction with higher agreement achieves a Pearson’s r = 0.999 (left-side of the figure), the case with poorest agreement achieves an r = 0.674 (right-side of the figure). Both raters have chosen the same AIF voxel in the best performance scenario. In the top-left part of Fig.  \ref{fig:localization} it can be seen that just a few voxels are activated in the displayed CT slice, having high activation values. The AIF voxel selected by the raters ($y_{r1,r2}(t)$) is also being activated by AIFNet, being the second highest value of $P_{vol}$. Mainly voxels belonging to the anterior cerebral artery are chosen. Besides, the AIF that our method predicts follows closely the raters' function, with no observable delays and with almost no differences in the curves’ shape. On the other hand, localization results from the worst Pearson's correlation case shows a different behaviour. Several voxels belonging to different arteries are enhanced by the network with an homogeneous activation distribution. The anterior cerebral artery and middle cerebral arteries are mainly selected. When assessing $y_{r1}(t)$ and $\hat{y}(t)$, it is noticeable that the low Pearson’s $r$ obtained is driven by the time shift between the functions (which is 4 seconds measured at the curve peaks). In this case, AIFNet outperforms rater $\#$1 by estimating a vascular function with high agreement in morphology, which occurs much earlier than the manually selected one. We consider the annotation of rater $\#$1 suboptimal, probably because the voxel was chosen from an artery branch already affected by the occlusion. However, our prediction follows more closely $y_{r2}(t)$ (Pearson's r = 0.980). There is no observable function delay between $y_{r2}(t)$ and $\hat{y}(t)$, though a slightly wider FWHM can be appreciated for $\hat{y}(t)$.

\subsection{Parameter Maps and Lesion Quantification}
\begin{figure}[!t]
\includegraphics[width=\textwidth]{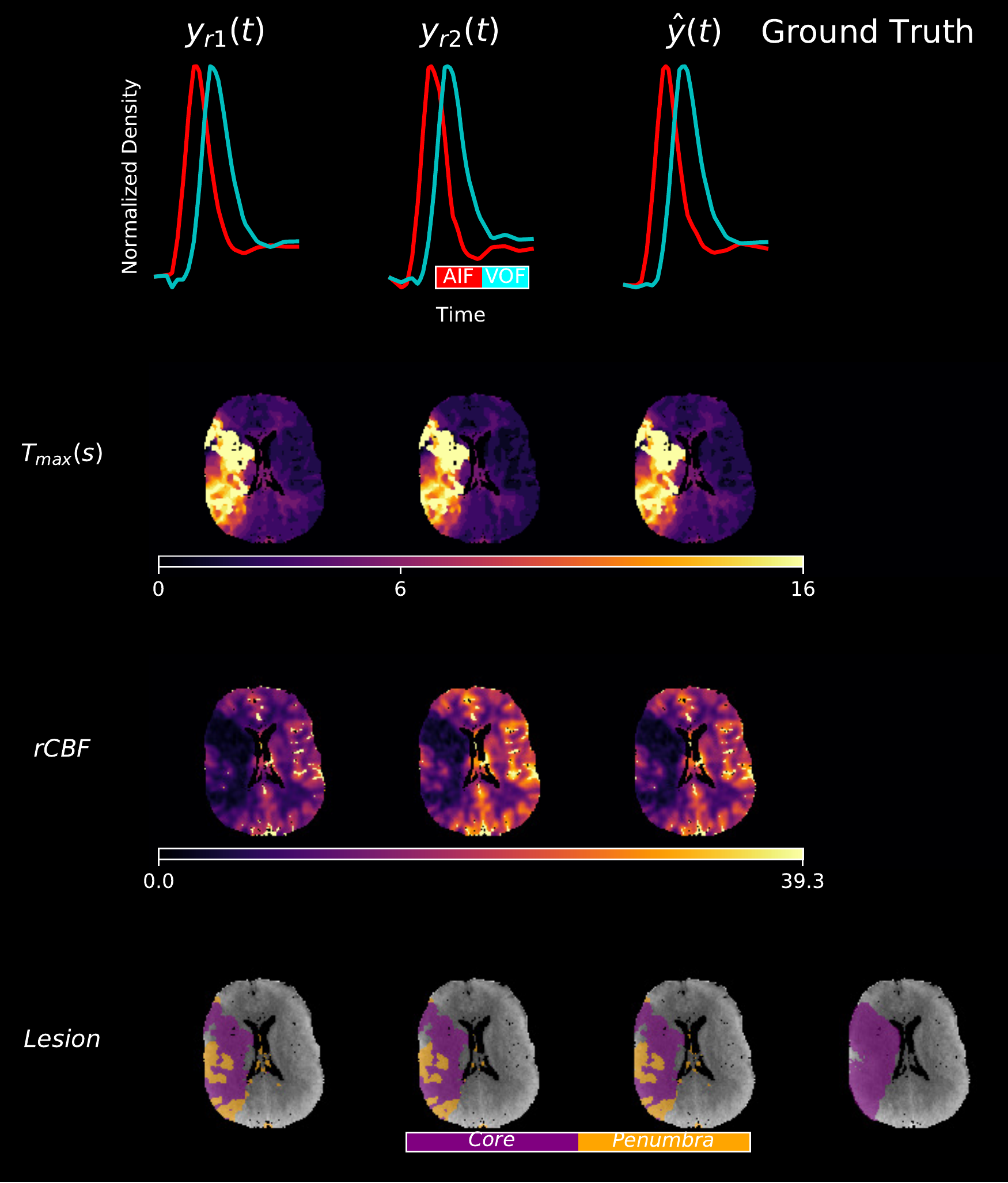}
\caption{Vascular functions, parameter maps and predicted lesions. The visualized ground truth masks correspond to the ISLES18 challenge delineations, which have been performed over DWI. Parameter maps and estimated lesion masks are obtained after deconvolving the CTP images with the annotated and predicted pair (AIF and VOF) of vascular functions. Estimated core lesions are obtained after thresholding the rCBF map at 38\% over the entire perfusion lesion (i.e. $T_{max}>6$ s). $y_{r1}(t)$: annotated pair of vascular functions by rater \#1; $y_{r2}(t)$: annotated pair of vascular functions by rater \#2; $\hat{y}(t)$: predicted pair of vascular functions by AIFNet; rCBF: relative cerebral-blood-flow map; Tmax: time to the maximum of the residue function map.}
\label{fig:quality_maps}
\end{figure}

%% brief description of quality
Vascular functions, parameter maps and lesion masks obtained for the raters and for AIFNet are shown in Fig. \ref{fig:quality_maps} for the median AIF performance in terms of Pearson correlation. There is a high correspondence between raters and AIFNet at all levels. All the perfusion lesions achieved very similar performance, with mainly tiny observable differences in the $penumbra$ estimations.

\begin{figure}[!t]
\hspace*{3.5cm} 
\includegraphics[scale=0.6]{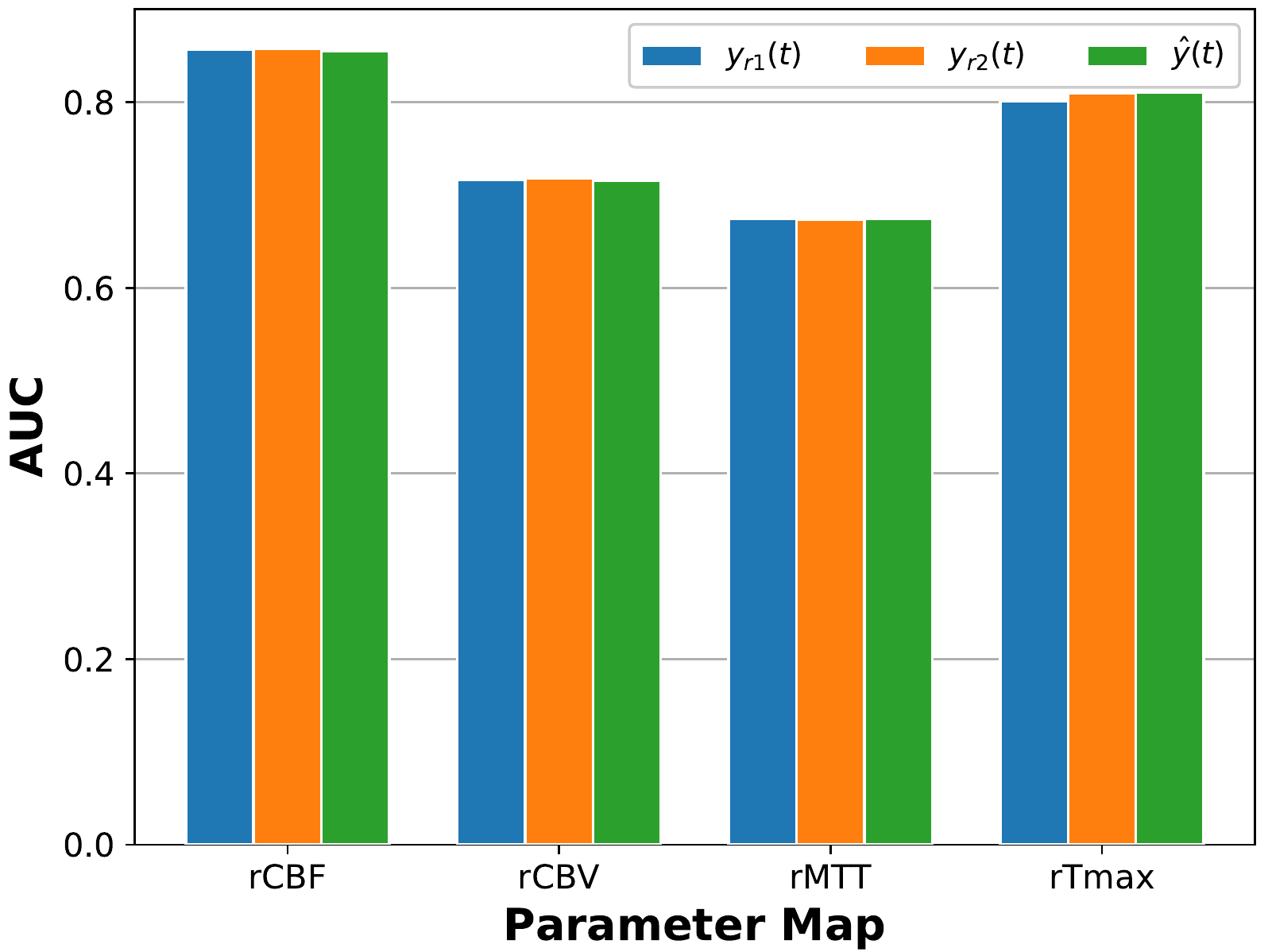}
\caption{ROC performance of the parameter maps for differentiating core and penumbra regions. Core lesions from the training ISLES18 challenge set are used as ground truth, which have been delineated over DWI. Parameter maps are obtained after deconvolving the CTP images with the annotated and predicted pair (AIF and VOF) of vascular functions. AUC: area under the curve. $y_{r1}(t)$: annotated pair of vascular functions by rater 1; $y_{r2}(t)$: annotated pair of vascular functions by rater 2; $\hat{y}(t)$: predicted pair of vascular functions by AIFNet; rCBF: relative cerebral-blood-flow; rCBV: relative cerebral-blood-volume; rMTT: relative mean-transit-time; rTmax: relative time to the maximum of the residue function.}
\label{fig:barplot}
\end{figure}

The parameter maps correlation between $y_{r1}(t)$ and $y_{r2}(t)$ and between $y_{r1}(t)$ and $\hat{y}(t)$ is 0.999 and 0.998 for rCBF, 0.999 and 0.999 for rCBV, 0.997 and 0.996 for rMTT and 0.862 and 0.833 for rTmax (Spearman correlation is used for rMTT and rTmax). When the parameters are assessed for the discrimination of the core lesion from the penumbra, the AUC values of Fig. \ref{fig:barplot} are obtained. There is a high agreement between raters and AIFNet, suggesting that the parameter maps that our method predicts have similar discriminant power for differentiating core and penumbra as the manually generated ones.  

In table \ref{tab_volumes} the brain core lesion quantification performance is shown for both the raters and for AIFNet. All metrics show a high agreement reaching inter-rater level. There are no statistically significant differences in Dice coefficients between raters (p-value = 0.91) nor in the AIFNet $vs$ raters comparison (p-value = 0.43 and p-value = 0.50 when comparing the core lesions obtained from $\hat{y}(t)$ with $y_{r1}(t)$ and $y_{r2}(t)$ respectively). When comparing the core lesion volumes against the DWI ground truths, both raters and AIFNet perform very similarly, sharing close error ranges. However, there is statistical significance in all volumetric comparisons of raters and AIFNet against the DWI ground truth (p-values $<$ 0.01, Wilcoxon test). 

The methods' comparison using parameter maps and lesion quantification show that the small AIF differences found with AIFNet (mainly in $T_{peak}$ and FWHM) are not strong enough to affect the deconvolution process. There are no consistent differences and no clear trend favoring one observer or AIFNet at vascular functions, parameter maps and core lesion quantification performance, suggesting our results reach inter-rater agreement.

\begin{table*}[!t]
\caption{\label{tab_volumes} Core lesion quantification performance (mean and (5th, 95th) percentiles) obtained after deconvolution of the CTP images with the annotated and predicted pair (AIF and VOF) of vascular functions. Core lesions from the training ISLES18 challenge set are used as ground truth, which have been delineated over DWI. Estimated core lesions are obtained after thresholding the rCBF map at 38\% over the entire perfusion lesion (i.e. $T_{max}<6$ s). $y_{r1}(t)$: annotated pair of vascular functions by rater 1; $y_{r2}(t)$: annotated pair of vascular functions by rater 2; $\hat{y}(t)$: predicted pair of vascular functions by AIFNet; r: Pearson's correlation coefficient.
}
\centering
\begin{tabular}{M{1.5cm}M{3.5cm}M{1.5cm}M{3.5cm}M{3.5cm}}
       &                 &               &              &               \\

       & \textbf{Dice (\%)}   & \textbf{r}    & \textbf{ Volume Error (ml)}                  & \textbf{ Absolute Volume Error (ml)}               \\
       \hline   
$y_{r1}(t)$ & 38.40 (2.70, 68.18) & 0.89  & 6.91 (-23.05, 44.69)  & 14.31 (0.62, 44.69) \\
$y_{r2}(t)$  & 38.42 (3.04, 68.31) & 0.89 & 6.64 (-21.74, 45.24)  & 14.22 (0.61, 45.24) \\
$\hat{y}(t)$ & 38.20 (2.37, 68.44) & 0.88 & 7.25 (-23.05, 44.69) & 14.51 (0.66, 45.67)\\
\hline

\end{tabular}
\end{table*}

\subsection{Ablation Analysis and Comparison with other CNNs}
Results for the ablation analysis are shown in Table \ref{tab_ablation}. Our experiments show that $K = 5$ convolutional layers are optimal for AIF prediction. The usage of less convolutional blocks leads not only to lower mean performance but also to higher variability. Besides, results do not improve when considering more than $K = 5$ convolutional layers. It is worth to point out the considerable improvement in robustness when problem-specific data augmentation is considered for training the models. Overall, a much higher 5th percentile is obtained rather than without data augmentation, showing better generalization over challenging cases. For VOF prediction $K$ = 2 convolutional layers are enough to estimate the function at inter-rater performance. Thus, less features are required for finding good VOF voxel candidates. These results are expected given the higher task difficulty for selecting AIF over VOF, as shown in Tables \ref{tab_aif} and \ref{tab_vof} where a better agreement between raters is shown for VOF than for AIF. 
 
The performance of the different CNNs is summarized in Table \ref{tab_ablation}. Our innovative end-to-end trained network is the most suitable approach among the investigated ones for finding vascular functions. On one hand, the comparison of AIFNet with the segmentation network not only shows that the approach performs much better, but also provides the advantage of requiring minimal data labels. Thus, the annotation of even a single voxel is enough for reaching inter-rater performance, turning the method easier and faster to deploy. On the other hand, the regression network reaches almost the same performance as AIFNet. The reason behind it could be that, unlike the segmentation network, AIFNet and the regression approach are directly optimized over the vascular function itself, allowing to better learn the curve profiles. Although the regression network shows overall good performance, it does not provide voxelwise arterial localization, making the model not interpretable for clinicians and hardly transferable to a clinical setting.

\begin{table}[!b]
\caption{\label{tab_ablation} Comparison with other approaches and ablation study. $K$ : Number of convolutional layers in the CNN. Given GPU memory constrains, the AIFNet experiment with $K =6$ is conducted with $2^{2+k}$ kernels per layer instead of $2^{3+k}$ (such as the first layer has 8 kernels and the sixth one 256). DA: Data augmentation; std: Standard deviation; Perc: Percentile.
}
\centering
\begin{tabular}{M{1cm}M{1cm}M{1cm}M{1cm}M{1cm}M{3cm}M{3cm}}
                 &   &   &    &                            &     &          \\
                 &   &   &    &                            &      &         \\
\multicolumn{5}{c}{\textbf{ Network }} &    \multicolumn{2}{c}{\textbf{Pearson’s r }}               \\    
& & & & &  Mean (std)                 & (5th, 95th Perc) \\
\hline

\multicolumn{5}{c}{AIFNet } &  & \\

\multicolumn{4}{c}{\textit{\# Layers (K) }} & & &               \\
 3 & 4 & 5 & 6&  DA &   &  \\
 \hline
 x &   &   & &   & 0.943 (0.133)              & 0.661-0.999   \\
& x &   & &   & 0.947 (0.107)              & 0.669-0.999   \\
 &   & x &  &  & 0.950 (0.088)              & 0.694-0.999   \\
  &   &  & x &  & 0.946 (0.094)              & 0.682-0.999   \\
&   & \textbf{x} & & \textbf{x} & \textbf{0.957 (0.057)}  &\textbf{ 0.870-0.999}\\
\hline 
\multicolumn{4}{c}{Regression AIFNet }& x& 0.949 (0.075) & 0.790-0.999 \\
\hline
\multicolumn{4}{c}{Segmentation AIFNet }&x & 0.796 (0.164) & 0.474-0.974 \\
\hline
\end{tabular}
\end{table}

\subsection{Limitations and future perspectives}
A limitation of this work is the core lesion ground truth used in the ISLES18 database, since currently there is no $gold$ standard for ischemic core. The acquisition delay between CTP and DWI imaging may introduce ischemic core modifications.  Moreover, the reperfusion therapy may also introduce some degree of mistmatch between the imaging modalities, since reversal of the DWI lesion may happen after reperfusion \citep{campbell2012infarct}. Consequently, a full correspondence between CTP and DWI core lesions is unlikely to happen. In our experiments, this  mismatch could explain the statistically significant differences found when comparing all CTP volumetric core predictions (from raters and AIFNet) against the delineated DWI ground truth.

As future perspectives for this work we could consider the validation of AIFNet over a larger database, as well as over other imaging modalities, such as perfusion MRI and PET images. Given the challenging task behind vascular estimation over CTP, we expect the method to be easy to adapt to images of better quality (such as MRI's). Exploring  whether the technique is generalizable to other organs and pathologies (such as brain tumors, myocardial infarction, etc.) also constitutes potential research lines. 
\section{Conclusions}
We have presented AIFNet, a new automatic method for vascular function estimation in brain perfusion imaging. It is developed and validated over the public ISLES18 database, which consists of stroke perfusion CT cases. For better reproducibility and direct comparison against future methods, we also provide both raters vascular annotations as supplementary material. To our knowledge, this is the first automatic method described in literature fully developed and validated over perfusion CT data. Most of the approaches previously described have been devised and tested over perfusion MRI instead.
For tackling the problem, we make use of a fully end-to-end trainable CNN, that is optimized for the prediction of vascular functions. We exploit prior knowledge by performing modality-specific data augmentation during the training stage. Our approach consistently differs from the previous ones, which mainly rely on clustering or statistical techniques providing a subset of suitable voxels to use for the selection of vascular functions. Additionally, most of these techniques require the definition of a decision rule (mainly a cutoff) for selecting the optimal voxel, a strategy that might be dataset-dependent and hence, requires parameters tuning. Unlike these methods, we present a non-heuristic function estimation strategy that combines multiple voxels information by means of a 3D probabilistic volume. AIFNet allows arterial localization and, hence, clinical interpretability. The method is easier to train and deploy compared to other approaches due to its architecture and due to the minimal voxel annotations required as ground truth (one single voxel per vascular function and per scan is enough to parametrize the network). As a consequence, the database labeling process is very fast. This is a clear advantage of AIFNet when compared against segmentation approaches, since the latter are more time consuming by requiring a vessel region annotation and multiple vascular functions checks. 
After validating AIFNet in the ISLES18 dataset, the method achieved results at an inter-rater variability level, being able to make predictions of vascular functions, parameter maps and core ischemic lesions with similar performance to human experts. Our results suggests that AIFNet could be implemented in clinical scenarios and hence, could potentially be included in future brain perfusion deconvolution software.

\section*{Disclosure}
EdlR and DR are co-inventors in technology related to this research; a patent application has been submitted and is pending. EdlR, DMS and DR are employees of ico\textbf{metrix}.

\section*{Acknowledgement}
This project received funding from the European Union's Horizon 2020 research and innovation program under the Marie Sklodowska-Curie grant agreement TRABIT No 765148. DR is supported by an innovation mandate of Flanders Innovation \& Entrepreneurship (VLAIO). BM received funding from the project ``Stroke treatment goes personalized: Gaining added diagnostic yield by computer-assisted treatment selection" (Deutsche Forschungsgemeinschaft (DFG) - Projekt number 326824585). 

\bibliography{mybibfile}

\begin{thebibliography}{56}
\expandafter\ifx\csname natexlab\endcsname\relax\def\natexlab#1{#1}\fi
\providecommand{\url}[1]{\texttt{#1}}
\providecommand{\href}[2]{#2}
\providecommand{\path}[1]{#1}
\providecommand{\DOIprefix}{doi:}
\providecommand{\ArXivprefix}{arXiv:}
\providecommand{\URLprefix}{URL: }
\providecommand{\Pubmedprefix}{pmid:}
\providecommand{\doi}[1]{\href{http://dx.doi.org/#1}{\path{#1}}}
\providecommand{\Pubmed}[1]{\href{pmid:#1}{\path{#1}}}
\providecommand{\bibinfo}[2]{#2}
\ifx\xfnm\relax \def\xfnm[#1]{\unskip,\space#1}\fi
%Type = Inproceedings
\bibitem[{Abulnaga and Rubin(2018)}]{abulnaga2018ischemic}
\bibinfo{author}{Abulnaga, S.M.}, \bibinfo{author}{Rubin, J.},
  \bibinfo{year}{2018}.
\newblock \bibinfo{title}{Ischemic stroke lesion segmentation in {CT} perfusion
  scans using pyramid pooling and focal loss}, in:
  \bibinfo{booktitle}{International MICCAI Brainlesion Workshop},
  \bibinfo{organization}{Springer}. pp. \bibinfo{pages}{352--363}.
%Type = Article
\bibitem[{Albers et~al.(2016)Albers, Goyal, Jahan, Bonafe, Diener, Levy,
  Pereira, Cognard, Cohen, Hacke et~al.}]{albers2016ischemic}
\bibinfo{author}{Albers, G.W.}, \bibinfo{author}{Goyal, M.},
  \bibinfo{author}{Jahan, R.}, \bibinfo{author}{Bonafe, A.},
  \bibinfo{author}{Diener, H.C.}, \bibinfo{author}{Levy, E.I.},
  \bibinfo{author}{Pereira, V.M.}, \bibinfo{author}{Cognard, C.},
  \bibinfo{author}{Cohen, D.J.}, \bibinfo{author}{Hacke, W.}, et~al.,
  \bibinfo{year}{2016}.
\newblock \bibinfo{title}{Ischemic core and hypoperfusion volumes predict
  infarct size in {SWIFT PRIME}}.
\newblock \bibinfo{journal}{Annals of neurology} \bibinfo{volume}{79},
  \bibinfo{pages}{76--89}.
%Type = Misc
\bibitem[{Bammer et~al.(2014)Bammer, Straka and Albers}]{bammer2014automated}
\bibinfo{author}{Bammer, R.}, \bibinfo{author}{Straka, M.},
  \bibinfo{author}{Albers, G.}, \bibinfo{year}{2014}.
\newblock \bibinfo{title}{Automated detection of arterial input function and/or
  venous output function voxels in medical imaging}.
\newblock \bibinfo{note}{US Patent 8,837,800}.
%Type = Inproceedings
\bibitem[{Bertels et~al.(2018)Bertels, Robben, Vandermeulen and
  Suetens}]{bertels2018contra}
\bibinfo{author}{Bertels, J.}, \bibinfo{author}{Robben, D.},
  \bibinfo{author}{Vandermeulen, D.}, \bibinfo{author}{Suetens, P.},
  \bibinfo{year}{2018}.
\newblock \bibinfo{title}{Contra-lateral information {CNN} for core lesion
  segmentation based on native {CTP} in acute stroke}, in:
  \bibinfo{booktitle}{International MICCAI Brainlesion Workshop},
  \bibinfo{organization}{Springer}. pp. \bibinfo{pages}{263--270}.
%Type = Article
\bibitem[{Calamante(2013)}]{calamante2013arterial}
\bibinfo{author}{Calamante, F.}, \bibinfo{year}{2013}.
\newblock \bibinfo{title}{Arterial input function in perfusion {MRI}: a
  comprehensive review}.
\newblock \bibinfo{journal}{Progress in nuclear magnetic resonance
  spectroscopy} \bibinfo{volume}{74}, \bibinfo{pages}{1--32}.
%Type = Article
\bibitem[{Calamante et~al.(2000)Calamante, Gadian and
  Connelly}]{calamante2000delay}
\bibinfo{author}{Calamante, F.}, \bibinfo{author}{Gadian, D.G.},
  \bibinfo{author}{Connelly, A.}, \bibinfo{year}{2000}.
\newblock \bibinfo{title}{Delay and dispersion effects in dynamic
  susceptibility contrast {MRI}: simulations using singular value
  decomposition}.
\newblock \bibinfo{journal}{Magnetic Resonance in Medicine: An Official Journal
  of the International Society for Magnetic Resonance in Medicine}
  \bibinfo{volume}{44}, \bibinfo{pages}{466--473}.
%Type = Article
\bibitem[{Calamante et~al.(2004)Calamante, M{\o}rup and
  Hansen}]{calamante2004defining}
\bibinfo{author}{Calamante, F.}, \bibinfo{author}{M{\o}rup, M.},
  \bibinfo{author}{Hansen, L.K.}, \bibinfo{year}{2004}.
\newblock \bibinfo{title}{Defining a local arterial input function for
  perfusion {MRI} using independent component analysis}.
\newblock \bibinfo{journal}{Magnetic Resonance in Medicine: An Official Journal
  of the International Society for Magnetic Resonance in Medicine}
  \bibinfo{volume}{52}, \bibinfo{pages}{789--797}.
%Type = Article
\bibitem[{Campbell and Parsons(2018)}]{campbell2018imaging}
\bibinfo{author}{Campbell, B.C.}, \bibinfo{author}{Parsons, M.W.},
  \bibinfo{year}{2018}.
\newblock \bibinfo{title}{Imaging selection for acute stroke intervention}.
\newblock \bibinfo{journal}{International Journal of Stroke}
  \bibinfo{volume}{13}, \bibinfo{pages}{554--567}.
%Type = Article
\bibitem[{Campbell et~al.(2012)Campbell, Purushotham, Christensen, Desmond,
  Nagakane, Parsons, Lansberg, Mlynash, Straka, De~Silva
  et~al.}]{campbell2012infarct}
\bibinfo{author}{Campbell, B.C.}, \bibinfo{author}{Purushotham, A.},
  \bibinfo{author}{Christensen, S.}, \bibinfo{author}{Desmond, P.M.},
  \bibinfo{author}{Nagakane, Y.}, \bibinfo{author}{Parsons, M.W.},
  \bibinfo{author}{Lansberg, M.G.}, \bibinfo{author}{Mlynash, M.},
  \bibinfo{author}{Straka, M.}, \bibinfo{author}{De~Silva, D.A.}, et~al.,
  \bibinfo{year}{2012}.
\newblock \bibinfo{title}{The infarct core is well represented by the acute
  diffusion lesion: sustained reversal is infrequent}.
\newblock \bibinfo{journal}{Journal of Cerebral Blood Flow \& Metabolism}
  \bibinfo{volume}{32}, \bibinfo{pages}{50--56}.
%Type = Article
\bibitem[{Cereda et~al.(2016)Cereda, Christensen, Campbell, Mishra, Mlynash,
  Levi, Straka, Wintermark, Bammer, Albers et~al.}]{cereda2016benchmarking}
\bibinfo{author}{Cereda, C.W.}, \bibinfo{author}{Christensen, S.},
  \bibinfo{author}{Campbell, B.C.}, \bibinfo{author}{Mishra, N.K.},
  \bibinfo{author}{Mlynash, M.}, \bibinfo{author}{Levi, C.},
  \bibinfo{author}{Straka, M.}, \bibinfo{author}{Wintermark, M.},
  \bibinfo{author}{Bammer, R.}, \bibinfo{author}{Albers, G.W.}, et~al.,
  \bibinfo{year}{2016}.
\newblock \bibinfo{title}{A benchmarking tool to evaluate computer tomography
  perfusion infarct core predictions against a {DWI} standard}.
\newblock \bibinfo{journal}{Journal of Cerebral Blood Flow \& Metabolism}
  \bibinfo{volume}{36}, \bibinfo{pages}{1780--1789}.
%Type = Article
\bibitem[{Cl{\`e}rigues et~al.(2019)Cl{\`e}rigues, Valverde, Bernal, Freixenet,
  Oliver and Llad{\'o}}]{clerigues2019acute}
\bibinfo{author}{Cl{\`e}rigues, A.}, \bibinfo{author}{Valverde, S.},
  \bibinfo{author}{Bernal, J.}, \bibinfo{author}{Freixenet, J.},
  \bibinfo{author}{Oliver, A.}, \bibinfo{author}{Llad{\'o}, X.},
  \bibinfo{year}{2019}.
\newblock \bibinfo{title}{Acute ischemic stroke lesion core segmentation in
  {CT} perfusion images using fully convolutional neural networks}.
\newblock \bibinfo{journal}{Computers in Biology and Medicine}
  \bibinfo{volume}{115}, \bibinfo{pages}{103487}.
%Type = Article
\bibitem[{Fan et~al.(2019)Fan, Bian, Wang, Wang, Yang, Ji and
  Kang}]{fan2019automatic}
\bibinfo{author}{Fan, S.}, \bibinfo{author}{Bian, Y.}, \bibinfo{author}{Wang,
  E.}, \bibinfo{author}{Wang, D.J.}, \bibinfo{author}{Yang, Q.},
  \bibinfo{author}{Ji, X.}, \bibinfo{author}{Kang, Y.}, \bibinfo{year}{2019}.
\newblock \bibinfo{title}{An automatic estimation of arterial input function
  based on multi-stream {3D CNN}}.
\newblock \bibinfo{journal}{Frontiers in neuroinformatics}
  \bibinfo{volume}{13}, \bibinfo{pages}{49}.
%Type = Article
\bibitem[{Fieselmann et~al.(2011)Fieselmann, Kowarschik, Ganguly, Hornegger and
  Fahrig}]{fieselmann2011deconvolution}
\bibinfo{author}{Fieselmann, A.}, \bibinfo{author}{Kowarschik, M.},
  \bibinfo{author}{Ganguly, A.}, \bibinfo{author}{Hornegger, J.},
  \bibinfo{author}{Fahrig, R.}, \bibinfo{year}{2011}.
\newblock \bibinfo{title}{Deconvolution-based {CT} and {MR} brain perfusion
  measurement: theoretical model revisited and practical implementation
  details}.
\newblock \bibinfo{journal}{Journal of Biomedical Imaging}
  \bibinfo{volume}{2011}, \bibinfo{pages}{14}.
%Type = Article
\bibitem[{Force(1989)}]{force1989stroke}
\bibinfo{author}{Force, W.T.}, \bibinfo{year}{1989}.
\newblock \bibinfo{title}{Stroke-1989. {R}ecommendations on stroke prevention,
  diagnosis, and therapy. {R}eport of the {WHO} {T}ask {F}orce on {S}troke and
  other {C}erebrovascular {D}isorders}.
\newblock \bibinfo{journal}{Stroke} \bibinfo{volume}{20},
  \bibinfo{pages}{1407--1431}.
%Type = Article
\bibitem[{Hinkle and Guanci(2007)}]{hinkle2007acute}
\bibinfo{author}{Hinkle, J.L.}, \bibinfo{author}{Guanci, M.M.},
  \bibinfo{year}{2007}.
\newblock \bibinfo{title}{Acute ischemic stroke review}.
\newblock \bibinfo{journal}{Journal of neuroscience nursing}
  \bibinfo{volume}{39}, \bibinfo{pages}{285--293}.
%Type = Article
\bibitem[{Kamnitsas et~al.(2017)Kamnitsas, Ledig, Newcombe, Simpson, Kane,
  Menon, Rueckert and Glocker}]{kamnitsas2017efficient}
\bibinfo{author}{Kamnitsas, K.}, \bibinfo{author}{Ledig, C.},
  \bibinfo{author}{Newcombe, V.F.}, \bibinfo{author}{Simpson, J.P.},
  \bibinfo{author}{Kane, A.D.}, \bibinfo{author}{Menon, D.K.},
  \bibinfo{author}{Rueckert, D.}, \bibinfo{author}{Glocker, B.},
  \bibinfo{year}{2017}.
\newblock \bibinfo{title}{Efficient multi-scale {3D CNN} with fully connected
  {CRF} for accurate brain lesion segmentation}.
\newblock \bibinfo{journal}{Medical image analysis} \bibinfo{volume}{36},
  \bibinfo{pages}{61--78}.
%Type = Article
\bibitem[{Kealey et~al.(2004)Kealey, Loving, Delong and
  Eastwood}]{kealey2004user}
\bibinfo{author}{Kealey, S.M.}, \bibinfo{author}{Loving, V.A.},
  \bibinfo{author}{Delong, D.M.}, \bibinfo{author}{Eastwood, J.D.},
  \bibinfo{year}{2004}.
\newblock \bibinfo{title}{User-defined vascular input function curves:
  influence on mean perfusion parameter values and signal-to-noise ratio}.
\newblock \bibinfo{journal}{Radiology} \bibinfo{volume}{231},
  \bibinfo{pages}{587--593}.
%Type = Article
\bibitem[{Khatri et~al.(2014)Khatri, Yeatts, Mazighi, Broderick, Liebeskind,
  Demchuk, Amarenco, Carrozzella, Spilker, Foster et~al.}]{khatri2014time}
\bibinfo{author}{Khatri, P.}, \bibinfo{author}{Yeatts, S.D.},
  \bibinfo{author}{Mazighi, M.}, \bibinfo{author}{Broderick, J.P.},
  \bibinfo{author}{Liebeskind, D.S.}, \bibinfo{author}{Demchuk, A.M.},
  \bibinfo{author}{Amarenco, P.}, \bibinfo{author}{Carrozzella, J.},
  \bibinfo{author}{Spilker, J.}, \bibinfo{author}{Foster, L.D.}, et~al.,
  \bibinfo{year}{2014}.
\newblock \bibinfo{title}{Time to angiographic reperfusion and clinical outcome
  after acute ischaemic stroke: an analysis of data from the {I}nterventional
  {M}anagement of {S}troke {(IMS III)} phase 3 trial}.
\newblock \bibinfo{journal}{The Lancet Neurology} \bibinfo{volume}{13},
  \bibinfo{pages}{567--574}.
%Type = Article
\bibitem[{Kistler et~al.(2013)Kistler, Bonaretti, Pfahrer, Niklaus and
  B{\"u}chler}]{kistler2013virtual}
\bibinfo{author}{Kistler, M.}, \bibinfo{author}{Bonaretti, S.},
  \bibinfo{author}{Pfahrer, M.}, \bibinfo{author}{Niklaus, R.},
  \bibinfo{author}{B{\"u}chler, P.}, \bibinfo{year}{2013}.
\newblock \bibinfo{title}{The virtual skeleton database: an open access
  repository for biomedical research and collaboration}.
\newblock \bibinfo{journal}{Journal of medical Internet research}
  \bibinfo{volume}{15}, \bibinfo{pages}{e245}.
%Type = Article
\bibitem[{Klotz and K{\"o}nig(1999)}]{klotz1999perfusion}
\bibinfo{author}{Klotz, E.}, \bibinfo{author}{K{\"o}nig, M.},
  \bibinfo{year}{1999}.
\newblock \bibinfo{title}{Perfusion measurements of the brain: using dynamic
  {CT} for the quantitative assessment of cerebral ischemia in acute stroke}.
\newblock \bibinfo{journal}{European journal of radiology}
  \bibinfo{volume}{30}, \bibinfo{pages}{170--184}.
%Type = Article
\bibitem[{Konstas et~al.(2009)Konstas, Goldmakher, Lee and
  Lev}]{konstas2009theoretic}
\bibinfo{author}{Konstas, A.}, \bibinfo{author}{Goldmakher, G.},
  \bibinfo{author}{Lee, T.Y.}, \bibinfo{author}{Lev, M.}, \bibinfo{year}{2009}.
\newblock \bibinfo{title}{Theoretic basis and technical implementations of {CT}
  perfusion in acute ischemic stroke, part 1: theoretic basis}.
\newblock \bibinfo{journal}{American Journal of Neuroradiology}
  \bibinfo{volume}{30}, \bibinfo{pages}{662--668}.
%Type = Article
\bibitem[{Kosior and Frayne(2007)}]{kosior2007perftool}
\bibinfo{author}{Kosior, J.C.}, \bibinfo{author}{Frayne, R.},
  \bibinfo{year}{2007}.
\newblock \bibinfo{title}{Perftool: A software platform for investigating
  bolus-tracking perfusion imaging quantification strategies}.
\newblock \bibinfo{journal}{Journal of Magnetic Resonance Imaging: An Official
  Journal of the International Society for Magnetic Resonance in Medicine}
  \bibinfo{volume}{25}, \bibinfo{pages}{653--659}.
%Type = Inproceedings
\bibitem[{Krizhevsky et~al.(2012)Krizhevsky, Sutskever and
  Hinton}]{krizhevsky2012imagenet}
\bibinfo{author}{Krizhevsky, A.}, \bibinfo{author}{Sutskever, I.},
  \bibinfo{author}{Hinton, G.E.}, \bibinfo{year}{2012}.
\newblock \bibinfo{title}{Imagenet classification with deep convolutional
  neural networks}, in: \bibinfo{booktitle}{Advances in neural information
  processing systems}, pp. \bibinfo{pages}{1097--1105}.
%Type = Article
\bibitem[{Kudo et~al.(2010)Kudo, Sasaki, Yamada, Momoshima, Utsunomiya, Shirato
  and Ogasawara}]{kudo2010differences}
\bibinfo{author}{Kudo, K.}, \bibinfo{author}{Sasaki, M.},
  \bibinfo{author}{Yamada, K.}, \bibinfo{author}{Momoshima, S.},
  \bibinfo{author}{Utsunomiya, H.}, \bibinfo{author}{Shirato, H.},
  \bibinfo{author}{Ogasawara, K.}, \bibinfo{year}{2010}.
\newblock \bibinfo{title}{Differences in {CT} perfusion maps generated by
  different commercial software: quantitative analysis by using identical
  source data of acute stroke patients}.
\newblock \bibinfo{journal}{Radiology} \bibinfo{volume}{254},
  \bibinfo{pages}{200--209}.
%Type = Article
\bibitem[{Lin et~al.(2016)Lin, Bivard, Krishnamurthy, Levi and
  Parsons}]{lin2016whole}
\bibinfo{author}{Lin, L.}, \bibinfo{author}{Bivard, A.},
  \bibinfo{author}{Krishnamurthy, V.}, \bibinfo{author}{Levi, C.R.},
  \bibinfo{author}{Parsons, M.W.}, \bibinfo{year}{2016}.
\newblock \bibinfo{title}{Whole-brain {CT} perfusion to quantify acute ischemic
  penumbra and core}.
\newblock \bibinfo{journal}{Radiology} \bibinfo{volume}{279},
  \bibinfo{pages}{876--887}.
%Type = Article
\bibitem[{Lorenz et~al.(2006)Lorenz, Benner, Chen, Lopez, Ay, Zhu, Menezes,
  Aronen, Karonen, Liu et~al.}]{lorenz2006automated}
\bibinfo{author}{Lorenz, C.}, \bibinfo{author}{Benner, T.},
  \bibinfo{author}{Chen, P.J.}, \bibinfo{author}{Lopez, C.J.},
  \bibinfo{author}{Ay, H.}, \bibinfo{author}{Zhu, M.W.},
  \bibinfo{author}{Menezes, N.M.}, \bibinfo{author}{Aronen, H.},
  \bibinfo{author}{Karonen, J.}, \bibinfo{author}{Liu, Y.}, et~al.,
  \bibinfo{year}{2006}.
\newblock \bibinfo{title}{Automated perfusion-weighted {MRI} using localized
  arterial input functions}.
\newblock \bibinfo{journal}{Journal of Magnetic Resonance Imaging: An Official
  Journal of the International Society for Magnetic Resonance in Medicine}
  \bibinfo{volume}{24}, \bibinfo{pages}{1133--1139}.
%Type = Article
\bibitem[{Maier et~al.(2017)Maier, Menze, von~der Gablentz, H{\"a}ni, Heinrich,
  Liebrand, Winzeck, Basit, Bentley, Chen et~al.}]{maier2017isles}
\bibinfo{author}{Maier, O.}, \bibinfo{author}{Menze, B.H.},
  \bibinfo{author}{von~der Gablentz, J.}, \bibinfo{author}{H{\"a}ni, L.},
  \bibinfo{author}{Heinrich, M.P.}, \bibinfo{author}{Liebrand, M.},
  \bibinfo{author}{Winzeck, S.}, \bibinfo{author}{Basit, A.},
  \bibinfo{author}{Bentley, P.}, \bibinfo{author}{Chen, L.}, et~al.,
  \bibinfo{year}{2017}.
\newblock \bibinfo{title}{Isles 2015-a public evaluation benchmark for ischemic
  stroke lesion segmentation from multispectral {MRI}}.
\newblock \bibinfo{journal}{Medical image analysis} \bibinfo{volume}{35},
  \bibinfo{pages}{250--269}.
%Type = Article
\bibitem[{McKinley et~al.(2018)McKinley, Hung, Wiest, Liebeskind and
  Scalzo}]{mckinley2018machine}
\bibinfo{author}{McKinley, R.}, \bibinfo{author}{Hung, F.},
  \bibinfo{author}{Wiest, R.}, \bibinfo{author}{Liebeskind, D.S.},
  \bibinfo{author}{Scalzo, F.}, \bibinfo{year}{2018}.
\newblock \bibinfo{title}{A machine learning approach to perfusion imaging with
  dynamic susceptibility contrast {MR}}.
\newblock \bibinfo{journal}{Frontiers in neurology} \bibinfo{volume}{9},
  \bibinfo{pages}{717}.
%Type = Article
\bibitem[{Meier et~al.(2019)Meier, Lux, Jung, Fischer, Gralla, Reyes, Wiest,
  McKinley and Kaesmacher}]{meier2019neural}
\bibinfo{author}{Meier, R.}, \bibinfo{author}{Lux, P.}, \bibinfo{author}{Jung,
  S.}, \bibinfo{author}{Fischer, U.}, \bibinfo{author}{Gralla, J.},
  \bibinfo{author}{Reyes, M.}, \bibinfo{author}{Wiest, R.},
  \bibinfo{author}{McKinley, R.}, \bibinfo{author}{Kaesmacher, J.},
  \bibinfo{year}{2019}.
\newblock \bibinfo{title}{Neural network--derived perfusion maps for the
  assessment of lesions in patients with acute ischemic stroke}.
\newblock \bibinfo{journal}{Radiology: artificial intelligence}
  \bibinfo{volume}{1}, \bibinfo{pages}{e190019}.
%Type = Article
\bibitem[{Mlynash et~al.(2005)Mlynash, Eyngorn, Bammer, Moseley and
  Tong}]{mlynash2005automated}
\bibinfo{author}{Mlynash, M.}, \bibinfo{author}{Eyngorn, I.},
  \bibinfo{author}{Bammer, R.}, \bibinfo{author}{Moseley, M.},
  \bibinfo{author}{Tong, D.C.}, \bibinfo{year}{2005}.
\newblock \bibinfo{title}{Automated method for generating the arterial input
  function on perfusion-weighted {MR} imaging: validation in patients with
  stroke}.
\newblock \bibinfo{journal}{American Journal of Neuroradiology}
  \bibinfo{volume}{26}, \bibinfo{pages}{1479--1486}.
%Type = Article
\bibitem[{Mouridsen et~al.(2006)Mouridsen, Christensen, Gyldensted and
  {\O}stergaard}]{mouridsen2006automatic}
\bibinfo{author}{Mouridsen, K.}, \bibinfo{author}{Christensen, S.},
  \bibinfo{author}{Gyldensted, L.}, \bibinfo{author}{{\O}stergaard, L.},
  \bibinfo{year}{2006}.
\newblock \bibinfo{title}{Automatic selection of arterial input function using
  cluster analysis}.
\newblock \bibinfo{journal}{Magnetic Resonance in Medicine: An Official Journal
  of the International Society for Magnetic Resonance in Medicine}
  \bibinfo{volume}{55}, \bibinfo{pages}{524--531}.
%Type = Article
\bibitem[{Murase et~al.(2001)Murase, Kikuchi, Miki, Shimizu and
  Ikezoe}]{murase2001determination}
\bibinfo{author}{Murase, K.}, \bibinfo{author}{Kikuchi, K.},
  \bibinfo{author}{Miki, H.}, \bibinfo{author}{Shimizu, T.},
  \bibinfo{author}{Ikezoe, J.}, \bibinfo{year}{2001}.
\newblock \bibinfo{title}{Determination of arterial input function using fuzzy
  clustering for quantification of cerebral blood flow with dynamic
  susceptibility contrast-enhanced {MR} imaging}.
\newblock \bibinfo{journal}{Journal of Magnetic Resonance Imaging: An Official
  Journal of the International Society for Magnetic Resonance in Medicine}
  \bibinfo{volume}{13}, \bibinfo{pages}{797--806}.
%Type = Article
\bibitem[{Murphy et~al.(2007)Murphy, Chen and Lee}]{murphy2007serial}
\bibinfo{author}{Murphy, B.}, \bibinfo{author}{Chen, X.}, \bibinfo{author}{Lee,
  T.Y.}, \bibinfo{year}{2007}.
\newblock \bibinfo{title}{Serial changes in {CT} cerebral blood volume and flow
  after 4 hours of middle cerebral occlusion in an animal model of embolic
  cerebral ischemia}.
\newblock \bibinfo{journal}{American Journal of Neuroradiology}
  \bibinfo{volume}{28}, \bibinfo{pages}{743--749}.
%Type = Article
\bibitem[{{\O}stergaard et~al.(1996a){\O}stergaard, Sorensen, Kwong, Weisskoff,
  Gyldensted and Rosen}]{ostergaard1996highII}
\bibinfo{author}{{\O}stergaard, L.}, \bibinfo{author}{Sorensen, A.G.},
  \bibinfo{author}{Kwong, K.K.}, \bibinfo{author}{Weisskoff, R.M.},
  \bibinfo{author}{Gyldensted, C.}, \bibinfo{author}{Rosen, B.R.},
  \bibinfo{year}{1996}a.
\newblock \bibinfo{title}{High resolution measurement of cerebral blood flow
  using intravascular tracer bolus passages. {P}art {II}: {E}xperimental
  comparison and preliminary results}.
\newblock \bibinfo{journal}{Magnetic resonance in medicine}
  \bibinfo{volume}{36}, \bibinfo{pages}{726--736}.
%Type = Article
\bibitem[{{\O}stergaard et~al.(1996b){\O}stergaard, Weisskoff, Chesler,
  Gyldensted and Rosen}]{ostergaard1996highI}
\bibinfo{author}{{\O}stergaard, L.}, \bibinfo{author}{Weisskoff, R.M.},
  \bibinfo{author}{Chesler, D.A.}, \bibinfo{author}{Gyldensted, C.},
  \bibinfo{author}{Rosen, B.R.}, \bibinfo{year}{1996}b.
\newblock \bibinfo{title}{High resolution measurement of cerebral blood flow
  using intravascular tracer bolus passages. {P}art {I}: {M}athematical
  approach and statistical analysis}.
\newblock \bibinfo{journal}{Magnetic resonance in medicine}
  \bibinfo{volume}{36}, \bibinfo{pages}{715--725}.
%Type = Article
\bibitem[{Peruzzo et~al.(2011)Peruzzo, Bertoldo, Zanderigo and
  Cobelli}]{peruzzo2011automatic}
\bibinfo{author}{Peruzzo, D.}, \bibinfo{author}{Bertoldo, A.},
  \bibinfo{author}{Zanderigo, F.}, \bibinfo{author}{Cobelli, C.},
  \bibinfo{year}{2011}.
\newblock \bibinfo{title}{Automatic selection of arterial input function on
  dynamic contrast-enhanced {MR} images}.
\newblock \bibinfo{journal}{Computer methods and programs in biomedicine}
  \bibinfo{volume}{104}, \bibinfo{pages}{e148--e157}.
%Type = Article
\bibitem[{Rausch et~al.(2000)Rausch, Scheffler, Rudin and
  Rad{\"u}}]{rausch2000analysis}
\bibinfo{author}{Rausch, M.}, \bibinfo{author}{Scheffler, K.},
  \bibinfo{author}{Rudin, M.}, \bibinfo{author}{Rad{\"u}, E.},
  \bibinfo{year}{2000}.
\newblock \bibinfo{title}{Analysis of input functions from different arterial
  branches with gamma variate functions and cluster analysis for quantitative
  blood volume measurements}.
\newblock \bibinfo{journal}{Magnetic resonance imaging} \bibinfo{volume}{18},
  \bibinfo{pages}{1235--1243}.
%Type = Article
\bibitem[{Rempp et~al.(1994)Rempp, Brix, Wenz, Becker, G{\"u}ckel and
  Lorenz}]{rempp1994quantification}
\bibinfo{author}{Rempp, K.A.}, \bibinfo{author}{Brix, G.},
  \bibinfo{author}{Wenz, F.}, \bibinfo{author}{Becker, C.R.},
  \bibinfo{author}{G{\"u}ckel, F.}, \bibinfo{author}{Lorenz, W.J.},
  \bibinfo{year}{1994}.
\newblock \bibinfo{title}{Quantification of regional cerebral blood flow and
  volume with dynamic susceptibility contrast-enhanced {MR} imaging.}
\newblock \bibinfo{journal}{Radiology} \bibinfo{volume}{193},
  \bibinfo{pages}{637--641}.
%Type = Article
\bibitem[{Robben(2016)}]{robben2016image}
\bibinfo{author}{Robben, D.}, \bibinfo{year}{2016}.
\newblock \bibinfo{title}{Image-based quantification of cerebral vascular
  connectivity. {D}octoral dissertation. {R}etrieved from
  https://limo.libis.be/. {K}atholieke {U}niversiteit {L}euven, {L}euven,
  {B}elgium} .
%Type = Article
\bibitem[{Robben et~al.(2020)Robben, Boers, Marquering, Langezaal, Roos, van
  Oostenbrugge, van Zwam, Dippel, Majoie, van~der Lugt
  et~al.}]{robben2020prediction}
\bibinfo{author}{Robben, D.}, \bibinfo{author}{Boers, A.M.},
  \bibinfo{author}{Marquering, H.A.}, \bibinfo{author}{Langezaal, L.L.},
  \bibinfo{author}{Roos, Y.B.}, \bibinfo{author}{van Oostenbrugge, R.J.},
  \bibinfo{author}{van Zwam, W.H.}, \bibinfo{author}{Dippel, D.W.},
  \bibinfo{author}{Majoie, C.B.}, \bibinfo{author}{van~der Lugt, A.}, et~al.,
  \bibinfo{year}{2020}.
\newblock \bibinfo{title}{Prediction of final infarct volume from native {CT}
  perfusion and treatment parameters using deep learning}.
\newblock \bibinfo{journal}{Medical image analysis} \bibinfo{volume}{59},
  \bibinfo{pages}{101589}.
%Type = Inproceedings
\bibitem[{Robben and Suetens(2018)}]{robben2018perfusion}
\bibinfo{author}{Robben, D.}, \bibinfo{author}{Suetens, P.},
  \bibinfo{year}{2018}.
\newblock \bibinfo{title}{Perfusion parameter estimation using neural networks
  and data augmentation}, in: \bibinfo{booktitle}{International MICCAI
  Brainlesion Workshop}, \bibinfo{organization}{Springer}. pp.
  \bibinfo{pages}{439--446}.
%Type = Article
\bibitem[{Shi and Malik(2000)}]{shi2000normalized}
\bibinfo{author}{Shi, J.}, \bibinfo{author}{Malik, J.}, \bibinfo{year}{2000}.
\newblock \bibinfo{title}{Normalized cuts and image segmentation}.
\newblock \bibinfo{journal}{Departmental Papers (CIS)} , \bibinfo{pages}{107}.
%Type = Misc
\bibitem[{Shi et~al.(2014a)Shi, Defeng and Heng}]{shi2014systems}
\bibinfo{author}{Shi, L.}, \bibinfo{author}{Defeng, W.}, \bibinfo{author}{Heng,
  P.A.}, \bibinfo{year}{2014}a.
\newblock \bibinfo{title}{Systems and methods for detecting arterial input
  function {(AIF)}}.
\newblock \bibinfo{note}{US Patent 8,798,349}.
%Type = Article
\bibitem[{Shi et~al.(2014b)Shi, Wang, Liu, Fang, Wang, Huang, King, Heng and
  Ahuja}]{shi2014automatic}
\bibinfo{author}{Shi, L.}, \bibinfo{author}{Wang, D.}, \bibinfo{author}{Liu,
  W.}, \bibinfo{author}{Fang, K.}, \bibinfo{author}{Wang, Y.X.J.},
  \bibinfo{author}{Huang, W.}, \bibinfo{author}{King, A.D.},
  \bibinfo{author}{Heng, P.A.}, \bibinfo{author}{Ahuja, A.T.},
  \bibinfo{year}{2014}b.
\newblock \bibinfo{title}{Automatic detection of arterial input function in
  dynamic contrast enhanced {MRI} based on affinity propagation clustering}.
\newblock \bibinfo{journal}{Journal of Magnetic Resonance Imaging}
  \bibinfo{volume}{39}, \bibinfo{pages}{1327--1337}.
%Type = Inproceedings
\bibitem[{Song and Huang(2018)}]{song2018integrated}
\bibinfo{author}{Song, T.}, \bibinfo{author}{Huang, N.}, \bibinfo{year}{2018}.
\newblock \bibinfo{title}{Integrated extractor, generator and segmentor for
  ischemic stroke lesion segmentation}, in: \bibinfo{booktitle}{International
  MICCAI Brainlesion Workshop}, \bibinfo{organization}{Springer}. pp.
  \bibinfo{pages}{310--318}.
%Type = Article
\bibitem[{Sourbron et~al.(2007)Sourbron, Luypaert, Morhard, Seelos, Reiser and
  Peller}]{sourbron2007deconvolution}
\bibinfo{author}{Sourbron, S.}, \bibinfo{author}{Luypaert, R.},
  \bibinfo{author}{Morhard, D.}, \bibinfo{author}{Seelos, K.},
  \bibinfo{author}{Reiser, M.}, \bibinfo{author}{Peller, M.},
  \bibinfo{year}{2007}.
\newblock \bibinfo{title}{Deconvolution of bolus-tracking data: a comparison of
  discretization methods}.
\newblock \bibinfo{journal}{Physics in Medicine \& Biology}
  \bibinfo{volume}{52}, \bibinfo{pages}{6761}.
%Type = Article
\bibitem[{Thijs et~al.(2004)Thijs, Somford, Bammer, Robberecht, Moseley and
  Albers}]{thijs2004influence}
\bibinfo{author}{Thijs, V.N.}, \bibinfo{author}{Somford, D.M.},
  \bibinfo{author}{Bammer, R.}, \bibinfo{author}{Robberecht, W.},
  \bibinfo{author}{Moseley, M.E.}, \bibinfo{author}{Albers, G.W.},
  \bibinfo{year}{2004}.
\newblock \bibinfo{title}{Influence of arterial input function on hypoperfusion
  volumes measured with perfusion-weighted imaging}.
\newblock \bibinfo{journal}{Stroke} \bibinfo{volume}{35},
  \bibinfo{pages}{94--98}.
%Type = Article
\bibitem[{Trialists’~Collaboration et~al.(2013)}]{trialists2013organised}
\bibinfo{author}{Trialists’~Collaboration, S.U.}, et~al.,
  \bibinfo{year}{2013}.
\newblock \bibinfo{title}{Organised inpatient (stroke unit) care for stroke}.
\newblock \bibinfo{journal}{Cochrane database syst rev} \bibinfo{volume}{9}.
%Type = Article
\bibitem[{Ulas et~al.(2018a)Ulas, Das, Thrippleton, Hern{\'a}ndez, Armitage,
  Makin, Wardlaw and Menze}]{ulas2018convolutional}
\bibinfo{author}{Ulas, C.}, \bibinfo{author}{Das, D.},
  \bibinfo{author}{Thrippleton, M.J.}, \bibinfo{author}{Hern{\'a}ndez,
  M.d.C.V.}, \bibinfo{author}{Armitage, P.A.}, \bibinfo{author}{Makin, S.D.},
  \bibinfo{author}{Wardlaw, J.M.}, \bibinfo{author}{Menze, B.H.},
  \bibinfo{year}{2018}a.
\newblock \bibinfo{title}{Convolutional neural networks for direct inference of
  pharmacokinetic parameters: Application to stroke dynamic contrast-enhanced
  {MRI}}.
\newblock \bibinfo{journal}{Frontiers in neurology} \bibinfo{volume}{9}.
%Type = Inproceedings
\bibitem[{Ulas et~al.(2018b)Ulas, Tetteh, Thrippleton, Armitage, Makin,
  Wardlaw, Davies and Menze}]{ulas2018direct}
\bibinfo{author}{Ulas, C.}, \bibinfo{author}{Tetteh, G.},
  \bibinfo{author}{Thrippleton, M.J.}, \bibinfo{author}{Armitage, P.A.},
  \bibinfo{author}{Makin, S.D.}, \bibinfo{author}{Wardlaw, J.M.},
  \bibinfo{author}{Davies, M.E.}, \bibinfo{author}{Menze, B.H.},
  \bibinfo{year}{2018}b.
\newblock \bibinfo{title}{Direct estimation of pharmacokinetic parameters from
  {DCE-MRI} using deep {CNN} with forward physical model loss}, in:
  \bibinfo{booktitle}{International Conference on Medical Image Computing and
  Computer-Assisted Intervention}, \bibinfo{organization}{Springer}. pp.
  \bibinfo{pages}{39--47}.
%Type = Article
\bibitem[{Vagal et~al.(2019)Vagal, Wintermark, Nael, Bivard, Parsons, Grossman
  and Khatri}]{vagal2019automated}
\bibinfo{author}{Vagal, A.}, \bibinfo{author}{Wintermark, M.},
  \bibinfo{author}{Nael, K.}, \bibinfo{author}{Bivard, A.},
  \bibinfo{author}{Parsons, M.}, \bibinfo{author}{Grossman, A.W.},
  \bibinfo{author}{Khatri, P.}, \bibinfo{year}{2019}.
\newblock \bibinfo{title}{Automated {CT} perfusion imaging for acute ischemic
  stroke: pearls and pitfalls for real-world use}.
\newblock \bibinfo{journal}{Neurology} \bibinfo{volume}{93},
  \bibinfo{pages}{888--898}.
%Type = Article
\bibitem[{Wang et~al.(2020)Wang, Song, Dong, Cui, Huang and
  Zhang}]{wang2020automatic}
\bibinfo{author}{Wang, G.}, \bibinfo{author}{Song, T.}, \bibinfo{author}{Dong,
  Q.}, \bibinfo{author}{Cui, M.}, \bibinfo{author}{Huang, N.},
  \bibinfo{author}{Zhang, S.}, \bibinfo{year}{2020}.
\newblock \bibinfo{title}{Automatic ischemic stroke lesion segmentation from
  computed tomography perfusion images by image synthesis and attention-based
  deep neural networks}.
\newblock \bibinfo{journal}{Medical Image Analysis} , \bibinfo{pages}{101787}.
%Type = Article
\bibitem[{Winder et~al.(2020)Winder, d’Esterre, Menon, Fiehler and
  Forkert}]{winder2020automatic}
\bibinfo{author}{Winder, A.}, \bibinfo{author}{d’Esterre, C.D.},
  \bibinfo{author}{Menon, B.K.}, \bibinfo{author}{Fiehler, J.},
  \bibinfo{author}{Forkert, N.D.}, \bibinfo{year}{2020}.
\newblock \bibinfo{title}{Automatic arterial input function selection in {CT}
  and {MR} perfusion datasets using deep convolutional neural networks}.
\newblock \bibinfo{journal}{Medical Physics} .
%Type = Article
\bibitem[{Wu et~al.(2003a)Wu, {\O}stergaard, Koroshetz, Schwamm, O'Donnell,
  Schaefer, Rosen, Weisskoff and Sorensen}]{wu2003effects}
\bibinfo{author}{Wu, O.}, \bibinfo{author}{{\O}stergaard, L.},
  \bibinfo{author}{Koroshetz, W.J.}, \bibinfo{author}{Schwamm, L.H.},
  \bibinfo{author}{O'Donnell, J.}, \bibinfo{author}{Schaefer, P.W.},
  \bibinfo{author}{Rosen, B.R.}, \bibinfo{author}{Weisskoff, R.M.},
  \bibinfo{author}{Sorensen, A.G.}, \bibinfo{year}{2003}a.
\newblock \bibinfo{title}{Effects of tracer arrival time on flow estimates in
  {MR} perfusion-weighted imaging}.
\newblock \bibinfo{journal}{Magnetic Resonance in Medicine: An Official Journal
  of the International Society for Magnetic Resonance in Medicine}
  \bibinfo{volume}{50}, \bibinfo{pages}{856--864}.
%Type = Article
\bibitem[{Wu et~al.(2003b)Wu, {\O}stergaard, Weisskoff, Benner, Rosen and
  Sorensen}]{wu2003tracer}
\bibinfo{author}{Wu, O.}, \bibinfo{author}{{\O}stergaard, L.},
  \bibinfo{author}{Weisskoff, R.M.}, \bibinfo{author}{Benner, T.},
  \bibinfo{author}{Rosen, B.R.}, \bibinfo{author}{Sorensen, A.G.},
  \bibinfo{year}{2003}b.
\newblock \bibinfo{title}{Tracer arrival timing-insensitive technique for
  estimating flow in {MR} perfusion-weighted imaging using singular value
  decomposition with a block-circulant deconvolution matrix}.
\newblock \bibinfo{journal}{Magnetic Resonance in Medicine: An Official Journal
  of the International Society for Magnetic Resonance in Medicine}
  \bibinfo{volume}{50}, \bibinfo{pages}{164--174}.
%Type = Article
\bibitem[{Yin et~al.(2015)Yin, Sun, Yang and Guo}]{yin2015automated}
\bibinfo{author}{Yin, J.}, \bibinfo{author}{Sun, H.}, \bibinfo{author}{Yang,
  J.}, \bibinfo{author}{Guo, Q.}, \bibinfo{year}{2015}.
\newblock \bibinfo{title}{Automated detection of the arterial input function
  using normalized cut clustering to determine cerebral perfusion by dynamic
  susceptibility contrast-magnetic resonance imaging}.
\newblock \bibinfo{journal}{Journal of Magnetic Resonance Imaging}
  \bibinfo{volume}{41}, \bibinfo{pages}{1071--1078}.

\end{thebibliography}

\end{document}